\documentclass[10pt,journal]{IEEEtran}
\usepackage{verbatim}
\usepackage{epsfig} % for postscript graphics files
\usepackage{epstopdf}
\usepackage{graphicx}
\usepackage{balance}
\usepackage{comment}
\usepackage{subfigure}
\usepackage{booktabs}
\usepackage{amsmath}
\usepackage{amssymb}
\usepackage{amsfonts}
\usepackage{bm}
\usepackage{colortbl}
\usepackage{tabularx}
\usepackage{color}
\usepackage{xspace}
\usepackage{graphicx}    % For importing graphics
\usepackage[vlined,algo2e]{algorithm2e}
\usepackage{algorithm}
\usepackage[noend]{algpseudocode}  
\usepackage{placeins}
\usepackage{multirow}
\usepackage{caption}
\usepackage{enumitem}
\usepackage{leading}

\DeclareMathOperator*{\argmin}{arg\,min}

\newcommand{\RN}[1]{%
  \textup{\uppercase\expandafter{\romannumeral#1}}%
}

\newcommand{\SystemName}{EnTrans\xspace}

\makeatletter
\def\@copyrightspace{\relax}
\makeatother

\begin{document}

\title{\SystemName: Leveraging Kinetic Energy Harvesting Signal for Transportation Mode Detection} 

\author{Guohao~Lan$^\dag$,~\IEEEmembership{Member,~IEEE,}
	Weitao~Xu$^\dag$$^\ast$,~\IEEEmembership{Member,~IEEE,}\thanks{$^\ast$Corresponding Author.}\thanks{$^\dag$Most of the work was done while the authors were with the University of New South Wales, Sydney, Australia.}	
	Dong~Ma,~\IEEEmembership{Student Member,~IEEE,}
	Sara~Khalifa,~\IEEEmembership{Member,~IEEE,} %\hfil\break
	Mahbub~Hassan,~\IEEEmembership{Senior Member,~IEEE,}
	Wen~Hu,~\IEEEmembership{Senior Member,~IEEE}% <-this % stops a space
	
	\IEEEcompsocitemizethanks{
	\IEEEcompsocthanksitem Guohao Lan is with the Department of Electrical and Computer Engineering, Duke University, Durham, NC 27708, USA.\protect \quad \protect(E-mail:guohao.lan@duke.edu)
	\IEEEcompsocthanksitem Weitao Xu is with College of Computer Science and Software Engineering, Shenzhen University, China. (Email:weitao.xu@szu.edu.cn)	
	\IEEEcompsocthanksitem Dong Ma, Mahbub Hassan, and Wen Hu are with the School of Computer Science and Engineering, University of New South Wales, Sydney, Australia.\protect \quad (E-mail: \{firstname.lastname\}@unsw.edu.au)
	\IEEEcompsocthanksitem Sara Khalifa is with Distributed Sensing Systems research group, Data61, CSIRO, Australia. 
	(Email:{sara.khalifa}@data61.csiro.au)
	\protect\\
	}
}

\IEEEtitleabstractindextext{
\begin{abstract}
Monitoring the daily transportation modes of an individual provides useful information in many application domains, such as urban design, real-time journey recommendation, as well as providing location-based services. In existing systems, accelerometer and GPS are the dominantly used signal sources for transportation context monitoring which drain out the limited battery life of the wearable devices very quickly. To resolve the high energy consumption issue, in this paper, we present \SystemName, which enables transportation mode detection by using only the kinetic energy harvester as an energy-efficient signal source. The proposed idea is based on the intuition that the vibrations experienced by the passenger during traveling with different transportation modes are distinctive. Thus, voltage signal generated by the energy harvesting devices should contain sufficient features to distinguish different transportation modes. We evaluate our system using over 28 hours of data, which is collected by eight individuals using a practical energy harvesting prototype. The evaluation results demonstrate that \SystemName is able to achieve an overall accuracy over 92\% in classifying five different modes while saving more than 34\% of the system power compared to conventional accelerometer-based approaches. 
\end{abstract}

}

\maketitle

\IEEEdisplaynontitleabstractindextext

\IEEEpeerreviewmaketitle

\section{Introduction}
\label{intro}

Thanks to the recent advancements in embedded technology, sensor-rich mobile and wearable devices have become an ubiquitous part of our life and enable many useful applications that improve our life quality~\cite{seneviratne2017survey}. Detecting and profiling the daily transportation mode of an individual is one kind of applications that aims to monitor our daily mobility patterns. Such mobility pattern is extremely useful in many domains. In mobile computing domain, the user's transportation pattern can be used to enable intelligent recommendation systems that provide personalized route recommendations~\cite{zheng2008understanding}. For instance, real-time public transportation information can be sent to the user if she is traveling by bus, or we can provide real-time traffic and road condition information to her if she is driving. Another potential use of the transportation mode inference is for the daily commuting survey. Conventional surveys are usually taken by telephone or paper questionnaires. Those strategies heavily rely on high-cost but low-accuracy human effort. In contrast, our daily used wearable devices can automatically detect and record our transportation mode. The data can be collected by edge or cloud server for future city planning research. Lastly, mobile applications could use the monitored transportation behavior to automatically calculate, for example, CO2-footprint or level of physical activity~\cite{hemminki2013accelerometer,froehlich2009ubigreen}. While there are many other benefits, the profiling of an individual's transportation habit usually involves long-term continuous sensing, processing, and data communication of the sensor data, which puts further pressure on the limited battery life of the wearable devices. Frequent battery recharging or replacement has, without doubt, became the major impediment to pervasive use of the system.

To extend the battery lifetime, researchers have been exploring kinetic energy harvesting (KEH) to scavenge energy released from human and machine motions to power the mobile devices~\cite{mitcheson2008energy,huang2015battery}. These research efforts are beginning to see the light of commercialization, examples include AMPY~\cite{AMPY} mobile motion-charger that harvests energy from our daily activities, and the SEQUENT smartwatch~\cite{SEQUENT} that can harvest energy from wrist motions. To meet the lightweight requirement, MicroGen~\cite{MicroGen} has introduced a semiconductor-based MEMS (Micro-Electro-Mechanical Systems energy harvesting chip with a form-factor of 1.0cm$^2$ which is compatible with today's wearable and mobile devices. More broadly, the developments in kinetic-powered devices are fostering new pervasive computing research that consider kinetic energy harvester not only as a source of power, but also as a motion sensor to detect a wide range of contexts~\cite{rao2014systems,khalifa2017harke,xiang2013powering,campbell2016perpetual}. This new sensing architecture utilizes the voltage output of energy harvester as an alternative to accelerometer, thereby saving energy that would have been consumed by the accelerometer in long-term monitoring.   

Following this trend, in this paper, we investigate the use of KEH signal for transportation mode detection. In particular, we aim to classify five widely used modalities that are available in our city: \textbf{\textit{bus}}, \textbf{\textit{train}}, \textbf{\textit{car}}, \textbf{\textit{ferry}}, and \textbf{\textit{light rail}}. The proposed idea is based on the intuition that vibrations experienced by subject when traveling with different transportation modes are different. Thus, voltage generated by the energy harvester should contain features to classify different transportation modes. However, unlike specialized motion sensor, e.g., accelerometer, that can capture very faint motions and provide signal with high quality and resolution, energy harvesters are insensitive to faint motions and the generated voltage signal are usually coarse. Moreover, different from accelerometers that provide signal of three axis, energy harvesters have only one-axis of output. In this regard, KEH based sensing usually exhibits a performance deficiency compared to accelerometer-based system, and thus, they are mainly used for coarse-grained classification~\cite{khalifa2017harke,campbell2016perpetual,mahbub2018computer} at the moment. %In addition, due to the similarities between different transportation modes, the accuracy achieved by state-of-the-art accelerometer-based transportation system is only 85\%~\cite{hemminki2013accelerometer,shin2015urban}, it becomes more challenging for us to improve and achieve an acceptable system accuracy by merely using the KEH signal. 

In this paper, to overcome the accuracy challenge of KEH-based transportation mode detection, we design a \textit{sparse representation based classification algorithm} to improve the system accuracy. To the best of our knowledge, this is the first comprehensive study in detecting transportation modes using the harvested voltage from the KEH devices. The main contributions of this paper are as follows:
\begin{itemize}
	\item[(1)] We propose a novel transportation detection system, \SystemName, which detects the transportation mode from the voltage signal generated by the KEH during the traveling of the subject. 
	\item[(2)] We develop a sparse representation based framework to classify different transportation modalities. The proposed framework leverages a dictionary learning based method to provide more compact representation of the activities, and thereby underpinning higher recognition performance in classification. Evaluation results indicate that our approach improves recognition accuracy by over $10\%$ compared to traditional classification algorithms such as k-Nearest Neighbors (kNN) and Support Vector Machine (SVM) that are widely used in conventional KEH-signal based activity recognition systems~\cite{khalifa2017harke,kalantarian2015monitoring}.
	\item[(3)] We design and build our prototype using off-the-shelf piezoelectric energy harvester. By using the prototype, we evaluate the proposed system with over 28 hours of transportation data collected from eight individuals. Our results show that, the proposed system can reach over $92\%$ accuracy.	
	\item[(4)] We conduct a detailed power consumption profile to demonstrate the superiority of \SystemName in energy saving. Our evaluation with four state-of-the-art low power accelerometers indicates that \SystemName outperforms the most power efficient accelerometer by reducing 34$\%$ of the overall system power consumption.	
\end{itemize}

Partial and preliminary results of this paper have appeared in our previous work~\cite{lan2016transportation}. In this paper we provide the following three major extensions to the conference version: (1) We have significantly extended the dataset from previously 3 hours with only three transportation modes to 28 hours with five different modes; and (2) a sparse representation based classification method has been designed to improve the classification accuracy by 10\% compared to the ones used in~\cite{lan2016transportation}; and (3) we conduct a power measurement study to demonstrate the benefits of the KEH-based system in reducing sensing induced power consumption. %The rest of the paper is structured as follows. We review the related work in Section~\ref{related} followed by a preview of kinetic energy harvesting in Section~\ref{section:vibration_energy_harvesting_background}. The details of system design and implementation are given in Section~\ref{system}, followed by the evaluation in Section~\ref{evaluation}. We conclude the paper in Section~\ref{sec:conclusion}.

\section{Related work}
\label{related}

Many intelligent mobile systems have been proposed to sense and monitor daily transportation mode of individuals by using a various of signal. This section summarizes existing works in intelligent transportation mode detection and kinetic energy harvesting based context sensing.

\subsection{Intelligent Transportation Mode Detection Systems}

Existing transportation mode detection systems can be broadly grouped into three categories givne the signal been used: Global Positioning System (GPS), WiFi/Cellular, and motion sensor based systems. In the following, we introduce related works in each of the three categories.

\textbf{GPS-based systems}. GPS is the most widely used information source for transportation mode detection~\cite{patterson2003inferring,zheng2008understanding,reddy2010using,stenneth2011transportation}, as it provides useful information including location and speed of movements. Zheng et al.~\cite{zheng2008understanding} use solely GPS to classify the modes among walking, driving, and riding bike. A novel set of features has been proposed to make the system robust to different traffic and weather conditions. In~\cite{patterson2003inferring}, the authors apply an unsupervised learning technique to detect the transportation mode of an individual. In addition to GPS signal, the historical trip information of the user is used to predict the traveler's destination as well as the trip purpose. Using GPS only as the detection signal will reduce system accuracy~\cite{reddy2010using,stenneth2011transportation}. To resolve this problem, Stenneth et al.~\cite{stenneth2011transportation} proposed the use of smartphone GPS together with the knowledge of underlying transportation network to achieve transportation mode detection. Similarly, Reddy et al.~\cite{reddy2010using} use GPS in conjunction with accelerometer to infer user's movements. However, GPS-based systems have several limitations. First, GPS sensor has high power consumption, which means that the battery of those mobile device will be drained out very quickly~\cite{zheng2008understanding,hemminki2013accelerometer}. Second, GPS requires unobstructed view to the satellites and works poor when the user is indoor or underground. Lastly, existing GPS-based solutions~\cite{reddy2010using,stenneth2011transportation} can only achieve very modest accuracy in recognizing different motorized transportation modes.

\textbf{WiFi/Cellular based systems}. In addition to GPS, the variations in radio signal have also been explored to infer user's movements. Sohn et al.~\cite{sohn2006mobility} identified basic human activities of standing, walking, and driving by using the GSM (Global System for Mobile communications) traces. Their system yields an overall accuracy of 85\%. Similarly, Muller et al.~\cite{muller2006practical} use the signal fluctuations in GSM cell tower to estimate whether a user is still, walking or in motorized transport. By leveraging the GSM cellular signal strength levels, they have trained a neural network model with 80\% of classification accuracy. Mun et al.~\cite{mun2008parsimonious} achieve an overall 88\% of accuracy by combining GSM and WiFi signal for the classification among dwelling, walking, and driving. Other than cellular network signal, Muthukrishnan et al.~\cite{muthukrishnan2007sensing} leverage the spectral characteristics of WiFi signal for human motion detection. Using the WiFi receiving signal strength of at a mobile phone, the proposed system can achieve 94\% of accuracy in classifying moving and still. Although the aforementioned systems are more energy-efficient compared to GPS-based ones, they are susceptible to the position and density of WiFi access points. As a result, their performance is not robust in outside rural areas and new environments.

\textbf{Motion sensor} is another widely used signal source~\cite{hemminki2013accelerometer,sankaran2014using,yu2014big,liang2017convolutional,fang2017learning}. The state-of-art work is proposed by Hemminki et al.~\cite{hemminki2013accelerometer} which utilizes the accelerometer on mobile phones to infer five transportation motions. A discrete hidden markov model has been introduced for the kinematic motion classification, while, an AdaBoost-based algorithm has been used for motorized mode detection. Their system can achieve 85\% of accuracy given a various of route and weather conditions. Kartik et al.~\cite{sankaran2014using} proposed the use of barometer for the same purpose. The main advantage of barometer is the position-independent characteristic, as it measures the variation of air pressure during movement instead of acceleration changes. However, as a trade-off it is not sensitive enough to detect speed and height changes which are important features to classify different motorized modes. Consequently, barometer can only be used for coarse-grained detection. More recently, efforts have been made in leveraging sensor-fusion to improve classification accuracy. Examples such as Yu et al.~\cite{yu2014big} merge the signal from accelerometer, gyroscope, and magnetometer to detect five transportation modes with 92.5\% of accuracy. Similarly, Fang et al.~\cite{fang2017learning} combine sensor-fusion together with deep learning technique to classify five modes with 95\% of accuracy. Although conventional sensor-based systems are promising in classification accuracy, they require multiple sensors as signal input. As reported in~\cite{yu2014big}, the use of sensor-fusion introduces 3mW of additional power consumption to the mobile devices.

\subsection{Kinetic Energy Harvesting based Context Sensing} 
Recent efforts in the literature are applying kinetic energy harvester as a power-free vibration sensor for sensing. An interesting fact for energy harvesting powered IoT devices is that the physical contexts the device is monitoring are usually related to the energy harvesting source of the harvester~\cite{rao2014systems}. For instance, for kinetic-powered wearable IoTs, Khalifa et al.~\cite{khalifa2017harke} have investigated the use of the AC voltage signal generated by KEH for wearable sensing. The underpin of the idea is that the kinetic energy transducer can be modeled as an inertial oscillating system when it operates in the inertial-force mode. This means that the harvester can serve as an inertial sensor to capture the external inertial force that applies to it. The proposed system can achieve $83\%$ of accuracy for classifying daily human activities. Following this trend, tremendous efforts have been made in leveraging the energy harvesting signal of KEH to achieve a wide range of human-centric sensing, such as health monitoring~\cite{kalantarian2015monitoring,lan2015estimating}, gait-based user authentication~\cite{weitao2016ndss}, and sports training~\cite{blank2016ball}. Kalantarian et al.~\cite{kalantarian2015monitoring} have designed a piezoelectric transducer-based wearable necklace for food-intake monitoring. The proposed system achieves over 80\% of accuracy in distinguishing food categories. In~\cite{blank2016ball}, Blank et al. proposed a ball impact localization system using a piezoelectric embedded table tennis racket. More recently, Xu et al.~\cite{weitao2016ndss,ma2018sehs} proposed an authentication system which utilizes the AC voltage signal to authenticate the user based on gait analysis. The proposed system can achieve an recognition accuracy of $95\%$ when five gait cycles are used. In addition to human-centric sensing, in~\cite{lan2017veh,lan2018hidden}, the authors proposed the use of KEH-transducer as an energy-efficient receiver for acoustic communication.

\section{Principle of Kinetic Energy Harvesting}
\label{section:vibration_energy_harvesting_background}
%In this section, we provide the basic background of utilizing kinetic energy harvesting for sensing.

\subsection{Kinetic Energy Harvesting}

\begin{figure}[]
	\centering
	%\vspace{-0.1in}
	\includegraphics[scale=0.62]{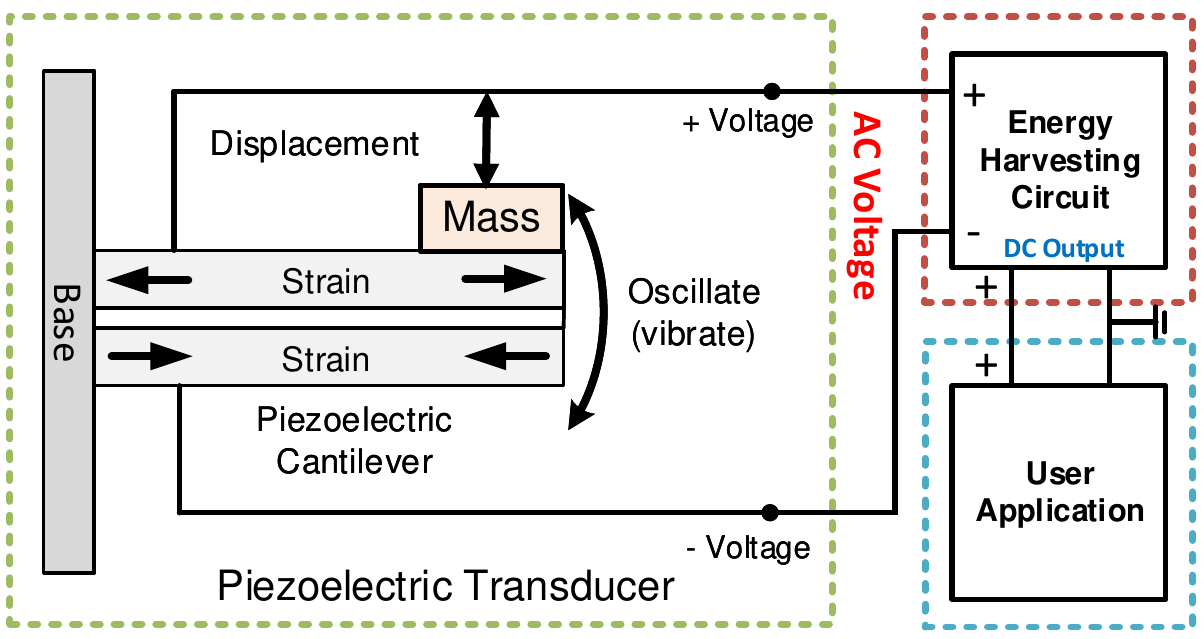}
	\caption{The principle of kinetic energy harvesting.}
	\label{Fig:PEH_background}
	\vspace{-0.1in}
\end{figure}

Kinetic energy harvesting refers to the process of scavenging kinetic energy released from human activity or ambient vibrations. The use of kinetic energy harvesting for self-powered IoT has been widely investigated in the literature~\cite{mitcheson2008energy,bhatti2016energy}. For KEH, there are three major energy transduction techniques that are widely used in the literature, namely, \textit{piezoelectric}, \textit{electromagnetic}, and \textit{electrostatic}. Among them, piezoelectric is the most favourable transduction mechanism for wearable IoTs, due to its simplicity and compatibility with MEMS (micro electrical mechanical system). Figure~\ref{Fig:PEH_background} exhibits a KEH system which utilizes a piezoelectric transducer to convert mechanical energy into electric AC voltage. The AC voltage is converted into DC output by the energy harvesting circuit to power external user applications (e.g., powering wearable devices or charging the batteries).

As shown, the piezoelectric transducer is usually modeled as an inertial oscillating system consisting of a cantilever beam attached with two piezoelectric outer-layers~\cite{mitcheson2008energy}. One end of the beam is fixed to the device, while the other is set free to oscillate (vibrate). When the piezoelectric cantilever is subjected to a mechanical stress, it expands on one side and contracts on the other. The induced piezoelectric effect will generate an alternating voltage (AC voltage) output as the beam oscillates around its neutral position. Theoretically, the AC output is proportional to the external mechanical stress/vibration applied, which indicates that the signal patterns of the AC voltage should reflect the external vibrational motions. In this paper, we built our proof-of-concept prototype based on the piezoelectric transducer. We use the AC voltage generated by the piezoelectric transducer as the signal source for transportation mode detection.

\section{System Design and Implementation} 
\label{system}

%\subsection{Intuition}
%The intuition behind the proposed system is that different transportation modes produce distinct vibrations, and the KEH is sensitive enough to capture such dynamics in the vibrations. As an example, Figure~\ref{fig:comparison} compares the voltage signal generated by the KEH from different transportation modalities in the time and frequency domain, respectively. We can see that different transportation modes exhibit distinguishable time domain and frequency domain features. Intuitively, walking and running produce higher voltage as they are heavily vibrant activities. In comparison, the generated voltage signal when the user is traveling by a vehicle, like car or train, is much moderate, since the user is in a stationary status when inside the vehicles. These features exhibit the possibility of using the KEH voltage signal for transportation modes detection.

\subsection{System Architecture}
\begin{figure}[]
	\centering
	\includegraphics[scale=0.25]{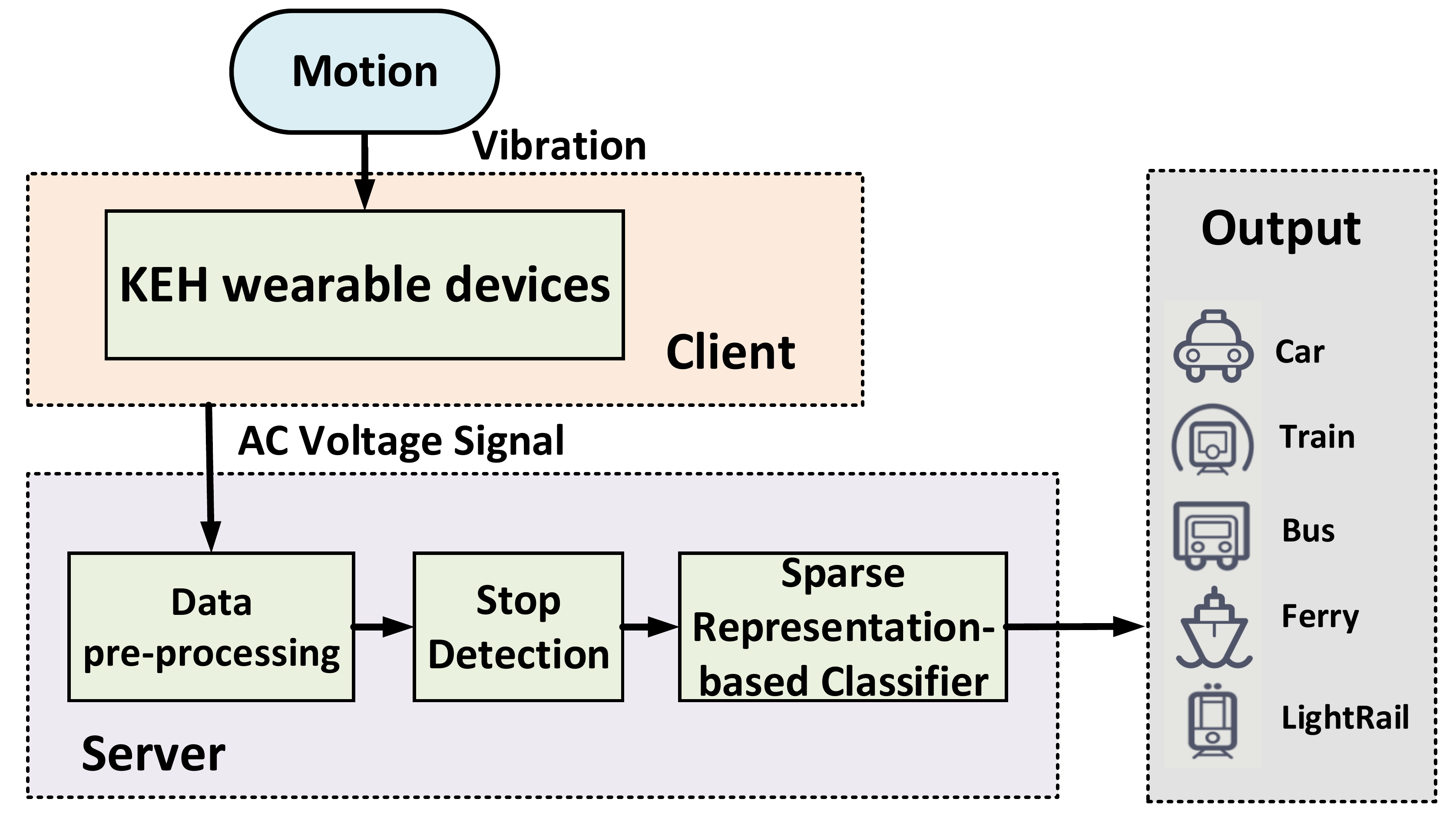}
	\caption{Overview of \SystemName system architecture.} 
	\label{Fig:Overview}
	\vspace{-0.15in}
\end{figure}

The overview of \SystemName is shown in Figure~\ref{Fig:Overview}, which consists of two parts: a mobile client and a server. The client is a wearable device which is carried by the subject during the traveling. It samples the voltage signal generated by the KEH and sends the collected data to the server where data processing and classification will be done. The proposed classification system is deployed at the server which could be a front-end edge device that has more powerful computation and energy resources while physically close to the user~\cite{shi2016edge}. In the transportation scenario, this edge device could be the computational unit deployed in the traveling vehicles or the ambient intelligent transportation infrastructures in the traveling route~\cite{datta2016integrating}. At the beginning of the system pipeline, the raw voltage signal collected by the KEH device is going through the data pre-processing for noise eliminating. In addition, a stop-detection algorithm is applied to identify and filter out the stop/pause segments from the signal profile (in Section~\ref{sec:signal_preprocessing}). As we will discuss later, without filtering out those signal, it will result in high classification error. After signal pre-processing and stop detection, the de-noised signal is feed into the a sparse representation-based classifier (Section~\ref{sec:mmc}) to determine the exact transportation mode among the five possibly modalities: \textbf{\textit{bus}}, \textbf{\textit{train}}, \textbf{\textit{car}}, \textbf{\textit{ferry}}, and \textbf{\textit{light rail}}. %Details of different system components will be given in the following section.

\subsection{Signal Pre-processing and Stop Detection}
\label{sec:signal_preprocessing}
As the energy harvester is not originally designed for sensing, the raw voltage signal generated by the energy harvester usually subjects to noise from the random and short-term vibrations due to user movements and external vibrations. In addition, hardware noise will also affect the signal quality. To eliminate the interference of noise, we first smooth the data using a moving average filter which is simple but effective for removing noise from time series. It smooths the data traces by replacing each raw data point with the average of the adjacent data points defined within the span (we use a span of 10 data points). Additionally, the stationary periods during the traveling, e.g., vehicle stops due to traffic light or arriving at the bus stop, will also introduce errors in the classification. For instance, the system can hardly classify the traveling mode of the subject when he/she is sitting in a stationary car or bus. We designed a stop detection algorithm to identify and filter out the stationary periods from the voltage data trace. Only the signal corresponds to the moving periods is kept and used as input for further analysis. 

For accelerometer-based system such as~\cite{stockx2014subwayps}, the stop of the vehicle is detected by comparing the average acceleration magnitude within a certain time window to a pre-defined threshold, or by using a probabilistic model of the acceleration magnitude to determine the status of the vehicle~\cite{thiagarajan2010cooperative}. In our system, the underlying idea is based on the fact that the AC voltage signal generated by the KEH device is fluctuating during traveling, whereas, it will be more stable within the stop/pause periods. Intuitively, this is because when the vehicle is stationary, the vibration applies to the KEH device is quite small and stable, and consequently, these features are reflected on the generated voltage signal. 

\begin{figure}[]
	\centering
	%\vspace{-0.1in}
	\includegraphics[scale=0.6]{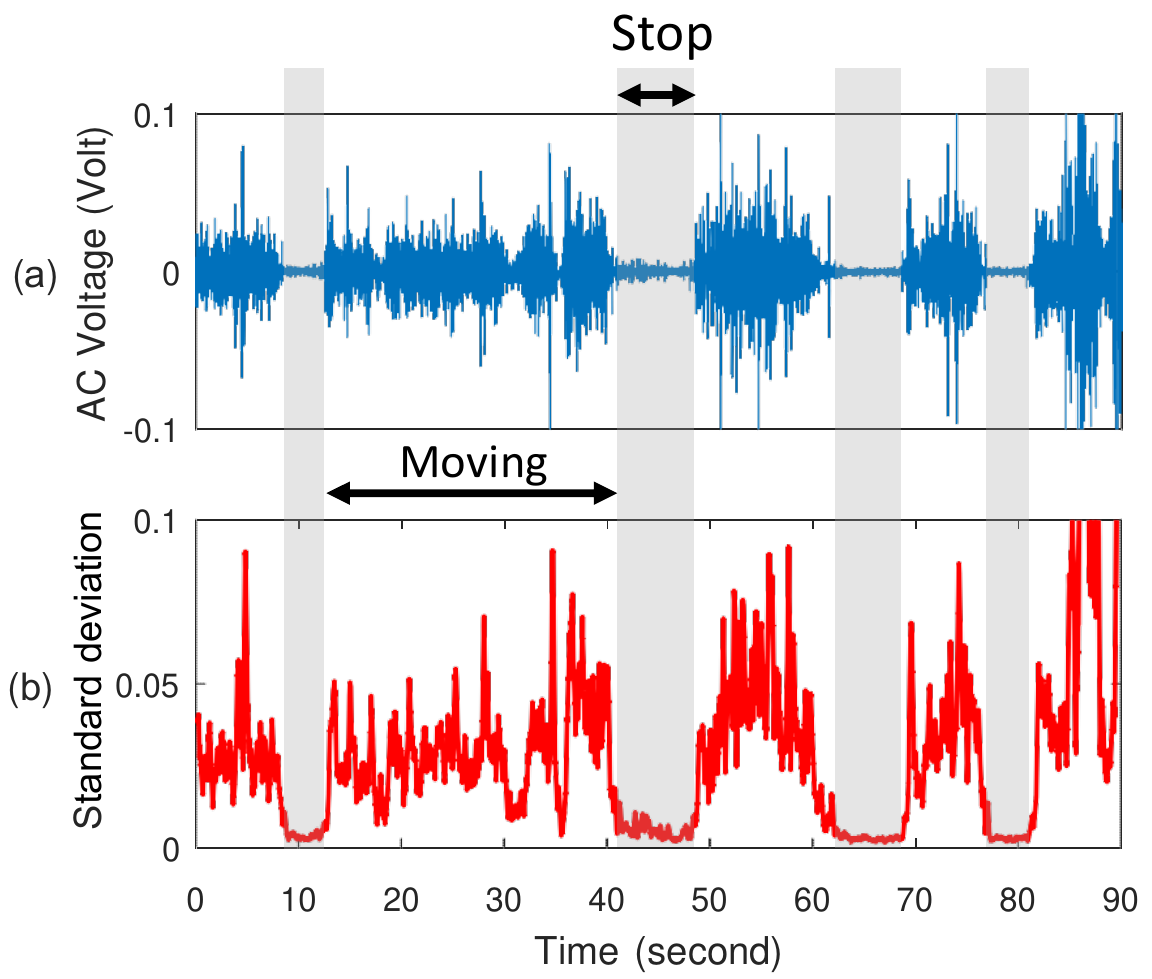}
	\caption{Example of stop detection: (a) AC voltage generated by KEH and (b) the standard deviation of the AC voltage signal during a car traveling.}
	\label{fig:stop_detection_example}
	\vspace{-0.2in}
\end{figure}

As an example, Figure~\ref{fig:stop_detection_example} exhibits the voltage signal collected during a car trip. We can clearly observe that there exists some periods in which the amplitude of the generated voltage signal are much lower and more stable than that in the other periods. These time periods are rightly corresponding to the stationary periods of the vehicle. Based on this observation, we design a thresholding algorithm to identify and filter out the samples that generated during these stationary periods. For a given voltage sample $v_t$ generated at time $t$, we calculate the standard deviation, $\sigma({t-k:t-1})$, using the previous $k$ samples that observed before $v_t$ over a one second window. The value of $k$ equals the sampling frequency used during the data sampling. For any voltage sample $v_t$ with $\sigma({t-k:t-1})$ smaller than a pre-defined threshold, we consider it as a voltage sample generated in the stationary period, and filtered it out from the signal trace. As shown in Figure~\ref{fig:stop_detection_example}(b), the standard deviation of the corresponding voltage signal within the stationary periods are much smaller than that of the moving periods. Therefore, we can effectively filter out those samples.

\subsection{Feature Selection and Classification}
\label{sec:mmc}

%In this study, we consider the following common urban transport modalities: \textbf{\textit{car}}, \textbf{\textit{bus}}, \textbf{\textit{train}}, \textbf{\textit{light rail}}, and \textbf{\textit{ferry}}. 

After de-noising and stop detection, the filtered KEH signal is feed into the sparse representation-based classifier to distinguish transportation modes. As discussed previously, the KEH transducer is insensitivity to the faint vibrations of the vehicles, which leads to high similarity in the generated voltage signal among different transportation modes. This poses a major challenge in the classification performance. To overcome the accuracy deficiency due to the use of KEH signal, in the following, we introduce our feature selection and the designed a sparse representation based classification framework. 

\subsubsection{\textbf{Feature Selection}}
\label{sec:feature_extraction}
Regardless of the classification technique been used, unambiguous classification is possible if and only if the signals of different transportation modes do not highly overlap in the feature space. For instance, if signals collected from the car and from the bus share similar feature values they cannot be unambiguously classified. Thus, feature selection is crucial for the classification system as it can help to select the most distinguishable features from a high-dimensional possible set. In the following, we introduce the features used in the motorized motion classifier, and the algorithm we used for feature selection.

\begin{table}[t]
	\centering
	\caption{Vibration-based features.}
	\label{vibarion_feature}
	\resizebox{3.4in}{!}{
	\begin{tabular}{|c|l|}
		\hline
		\textbf{Feature}        & \multicolumn{1}{c|}{\textbf{Description}}                                                                                               \\ \hline
		Mean of Peaks           & \begin{tabular}[c]{@{}l@{}}The average amplitude of the signal peaks in\\ the given sliding window.\end{tabular}                        \\ \hline
		Mean Peak Distance      & \begin{tabular}[c]{@{}l@{}}The average distance between two adjacent \\ signal peaks in the sliding window.\end{tabular}                \\ \hline
		Max Peak Distance       & \begin{tabular}[c]{@{}l@{}}The maximum distance between two adjacent \\ signal peaks.\end{tabular}                                      \\ \hline
		Max of Peaks            & \begin{tabular}[c]{@{}l@{}}The maximun amplitude of the signal peaks in \\ the sliding window.\end{tabular}                             \\ \hline
		Peak to Peak            & \begin{tabular}[c]{@{}l@{}}The amplitude difference between the maximum \\ and minimum signal peaks in the sliding window.\end{tabular} \\ \hline
		%Peak to Peak Difference & \begin{tabular}[c]{@{}l@{}}The amplitude different between tow adjacent \\ signal peaks.\end{tabular}                                   \\ \hline
	\end{tabular}
	}
	%\vspace{-0.1in}
\end{table}

\begin{table*}[t]
	\centering
	\caption{List of features and corresponding RMI.}
	\label{tab:feature}
	\small
	\resizebox{6.3in}{!}{
		\begin{tabular}{llc||llc} 
			\hline
			\textbf{Statistical Features}                 & \textbf{RMI}   &  \textbf{Selected} &                                    & \textbf{RMI}  & \textbf{Selected}\\ \hline
			Length                                        & 2.7\%  &   			& 1st and 3rd Quartile                   & 10.6\% &\checkmark  \\ \hline
			Min                                           & 15.4\% &\checkmark  & Skewness                           & 4.6\% & \\ \hline
			Mean                                          & 1.4\%  & 			& Kurtosis                           & 7.2\% &\\ \hline
			Median                                        & 0.7\%  & 			& Absolute area                      & 14.7\% &\checkmark \\ \hline
			Max                                           & 14.9\% &\checkmark  &\textbf{Vibration-based features}    &        & \\ \hline
			Standard deviation                                           & 14.3\% &\checkmark  & Mean of peaks                      & 16.4\% &\checkmark \\ \hline
			Root mean square                                           & 3.4\%  & 			& Mean peak distance     & 14.8\% &\checkmark \\ \hline
			Mean of absolute value                        & 13\%   &\checkmark  & Max peak distance         & 13.4\% &\checkmark \\ \hline
			Number of samples higher than threshold 1,2,3 & 19.4\% &\checkmark  & Max of peaks                       & 16.3\% &\checkmark \\ \hline
			spectral entropy                              & 9.4\%  & 			& Peak to peak                       & 12.4\% &\checkmark \\ \hline
			Spectrum peak position                        & 17.9\% &\checkmark  &             &  & \\ \hline
			FFT coefficients (1-50Hz)                     & 15.3\% &\checkmark  & \textbf{Frequency domain features} &  & \\ \hline
			\textbf{Time domain features}                 &        & 			& Two Dominant frequencies             & 16.8\% &\checkmark \\ \hline
			Range                                         & 2.5\%  &  			& Dominant frequency ratio           & 12.4\% &\checkmark \\ \hline
			Mean of absolute deviation                    & 6.4\%  & 			& Mean of power spectrum             & 4.5\%  & \\ \hline
			Number of datapoint cross mean                & 3.2\%  & 			& Total energy of spectrum           & 8.9\% &\checkmark  \\ \hline
			Coefficient variation                         & 1.6\%  & 			& Min of power spectrum              & 3.6\% &  \\ \hline
			Interquartile range                           & 7.4\%  &\checkmark  & Max of power spectrum              & 2.7\% & \\ \hline
		\end{tabular}
	}
\end{table*}

We segment the AC voltage signal using a sliding window of $T$ second with 10\% overlap, and extract two different types of features from the sliding window:
\setlength{\leftmargini}{1em}
\begin{itemize}
	\item{\textbf{Window-based features}}: include statistical, time-domain, and frequency-domain features. These features are widely used to effectively capture the general characteristics of time series data. For instance, Min, Max, and Standard deviation can capture the statistical characteristic of the samples in the time-domain. The 1st and 3rd Quartile measure the overall distribution of the signal samples in the classification window. Similar, frequency-domain features such as the Two Dominant Frequencies and the Total Energy of Spectrum capture the fundamental frequency and the energy of the signal, respectively. We suggest interested readers refer~\cite{bulling2014tutorial,liu1998feature} for details of feature calculation.
	
	\item{\textbf{Vibration-based features}}: in addition to window-based features, we also introduce the use of vibration-based features to better quantify and capture the patterns and severity of vibration in different transportation mode. For a given sliding window, signal peaks are identified using peak analysis. Then, the detected peaks are used to calculate different vibration-based features. Detailed descriptions of the features are given in Table~\ref{vibarion_feature}. For instance, the Peak-to-Peak feature indicates the maximum excursion of the signal wave. It captures the vibratory displacement in the signal which could be critical for the classification of different modes. For instance, we usually experience a higher vibration intensity when traveling by bus than by the light-rail. This intensity characteristic can be captured by the Peak-to-Peak feature for classification. Similarly, the Mean Peak Distance and Max Peak Distance features are used to capture how frequent the signal peaks appear in time.
\end{itemize}

%\subsection{Feature Selection}
In feature selection, we leverage the relative mutual information (RMI) as the metric, as it is able to measure how much information a feature contributes to the correct classification~\cite{schutze2008introduction}. It indicates the percentage of entropy that is removed from the classification problem when a particular feature is known~\cite{frank2013touchalytics}. The RMI of a given feature $F$ on the classification of transportation mode $C$ is defined as:
\begin{equation} \small
RMI(C,F)=\frac{H(C)-H(C|F)}{H(C)},
\label{eq:rmi}	
\end{equation}
where $H(C)$ is the marginal entropy of feature $C$, and $H(C|F)$ is the entropy of feature $C$ conditioned on feature $F$. The marginal entropy of a discrete random variable $C$ can be explicitly defined as: $H(C)=-\sum_{i=1}^{n}P(c_i)\log P(c_i)$, where $P(C)$ is the probability mass function of $C$. The conditional entropy of two discrete random variable $C$ and $F$ can be obtained by: $H(C|F)=-\sum_{i,j}^{}p(c_i,f_j)\log \frac{p(c_i,f_j)}{p(f_j)}$, where $p(c_i,f_j)$ is the probability when $C=c_i$ and $F=f_j$. For each feature, the value of RMI ranges from 0 to 1, whereas 0 indicates the feature carries no information about the classification problem $C$ while a value of 1 means the feature can determines $C$ correctly. In addition, we also consider the correlation between different features to select the most effective feature set. For example, a group of features with low RMI value may still be useful when combined together. We applied the Minimum Redundancy Maximum Relevance (mRMR) algorithm~\cite{ding2005minimum} to select the optimal set of features that share the highest amount of information with the classification results (i.e., transportation mode) while maintaining low redundancy with each other. 

The list of the selected features and their corresponding RMIs is shown in Table~\ref{tab:feature}. We can notice that, the most informative features are the statistical and vibration-based features. This is because the statistical features such as Min, Max, and Standard deviation can effectively capture the characteristics of high-frequency motions caused by the vehicle's engine and contact between the vehicle and surface~\cite{hemminki2013accelerometer}, while the vibration-based features can reflect the velocity and acceleration changes of the vehicle during the acceleration and breaking periods. For instance, we notice that the Max and Standard deviation when traveling by train are smaller than that when traveling by bus. This is simply because the user experiences heavier vibrations and velocity changes in the car, while it is more stable when staying on the train.

%\begin{figure}[]
%	\centering
%	%	\subfigure[Walking]{
%	%		\includegraphics[height=2.2in, width=1.62in]{figure/walk_fig.pdf}
%	%		\label{fig:walk}}
%%	\subfigure[Running]{
%%		\includegraphics[height=2.2in, width=1.62in]{figure/run_fig.pdf}
%%		\label{fig:run}}
%	\subfigure[Car]{
%		\includegraphics[height=2.2in, width=1.62in]{figure/car_fig.pdf}
%		\label{fig:car}}
%		\subfigure[Train]{
%			\includegraphics[height=2.2in, width=1.62in]{figure/train_fig.pdf}
%			\label{fig:train}
%		}
%	\caption{A comparison of the voltage signal from KEH for different transportation modalities.}
%	\label{fig:comparison}
%	\vspace{-0.1in}
%\end{figure}

\subsubsection{\textbf{Sparse Representation based Classification}}
\label{sec:src}
Sparse Representation based Classification (SRC) is an emerging classification method and has been successfully used in a variety of applications such as gait recognition~\cite{zhang2015accelerometer}, voice recognition~\cite{wei2013real}, and face recognition~\cite{weitao2016sensor}. The SRC solves a single-label classification problem, which aims to return the class that best matches a given test sample. Compared to traditional classifiers like SVM and KNN, SRC is more robust to environment noise due to the use of $\ell_1$ optimization. In the following, we describe how to build the training dictionary and obtain classification results in detail. 

\textbf{Step 1: Dictionary Construction and Sparse Representation}. 
To model transportation mode recognition as a sparse representation problem, we need to build a training dictionary $D$. Recent research shows that learning a dictionary by fitting a set of overcomplete basis vectors to a collection of training samples can generate more compact and informative representation from given data and achieve better recognition accuracy~\cite{aharon2006img}. We construct the training dictionary using dictionary learning technique. In particular, we first learn one single dictionary for each transportation mode, which is formed by a set of basis vectors learned by solving a sparse optimization problem. Then we construct the full dictionary by concatenating all the dictionaries together.

We refer the training and test samples as training vectors and test vectors. Suppose that we have $K$ classes indexed by $i = 1,...,K$. Class $i$ contains $n_i$ training examples that are denoted as $S_{i}=\{ s_1, s_2,..., s_{n_i}\}$. Each training example is assumed to be a column vector with $m$ elements (i.e., feature dimension). For class $k$, we aim to find an overcomplete dictionary matrix $D_{k} \in \mathbb{R}^{m \times K}$ over which a test vector has a sparse representation $X_{k}=\{x_1,x_2,...,x_{n_i}\}$. After that, the raw training examples $S_{i}$ can be linearly expressed by $n_{k}$ vectors in $D_{k}$ where $n_{k} \ll K$. The optimization problem of training a dictionary can be formulated as:
\begin{equation} \small
\argmin_{D^{k},X^{k}} \|S^{k}-D^{k}X^{k}\|^{2}_2 \quad \text{ subject to } \|x_{i}\|_0 \leq n_{k}.
\label{eq:ksvd}
\end{equation}
There are several dictionary learning algorithms that can be used to train a dictionary such as MOD~\cite{engan2000multi}, K-SVD~\cite{aharon2006img} and NMF~\cite{lee2001algorithms}. In this study, we choose K-SVD because it is efficient, flexible and works in conjunction with any pursuit algorithms. The K-SVD algorithm involves two stages: first, $D^k$ is fixed and the coefficient matrix $X^k$ is optimized by orthogonal matching pursuit (OMP) algorithm. Then the dictionary $D^k$ is updated using the calculated $X^{k}$. The process repeats until the stopping criterion (i.e., a fixed number of iterations) is achieved. The dictionary learning algorithm is detailed in Algorithm~\ref{alg:dictionarylwarning}.

\begin{algorithm}[t]
	%\SetAlgoLined
	\small
	\caption{Activity-Specific Dictionary Learning}
	\label{alg:dictionarylwarning}
	\begin{algorithmic}[0] %??????
		\State \textbf{Input:} Training samples $S=\{ s_1, s_2, s_3,..., s_{n}\}$, initial dictionary $D^0\in \mathbb{R}^{m \times K}$, target sparsity $\tau$; 
		\State \textbf{Output:} Dictionary $D$ and sparse coefficients matrix $X$;
		\State \textbf{Initialization:} set dictionary $D=D^0$; 
		\State \While {!= stopping criteria}{
			\State			$x_i=\argmin_{x} \|s_i-x\|^2_2 \quad \text{ subject to } \forall i \quad \|Dx\|_0 \leq \tau$;
			\State	\For{$j=1,...,m$} {
				\State	$J= \{$\textit{indices of the columns of} $X$ \textit{orthogonal to} $w_j$ ($j$\textit{-th} \textit{column of} $D)\}$;
				\State	$w_j=\argmin_w \|w^{T}D_J\|^2_2 \quad \text{ subject to } \|w\|_2=1$;
				\State	$D(j$\textit{-th row}$)=w^{T}_j$;}
			\State	}
	\end{algorithmic}
\end{algorithm}

A key idea behind SRC is to assemble all the training vectors from all classes into a {\sl dictionary} matrix $A$. Let $a_{i,j} \in \mathbb{R}^{m}$ denote the $j$-th training vector for the $i$-th class. The dictionary matrix $A \in \mathbb{R}^{m \times n}$ has the following form: 
%\begin{equation}
\begin{align}\small
A &=[A_{1}, A_{2}, \ldots, A_{i}] \\
&=[a_{1,1}, \ldots,a_{1,n_1}, \ldots,a_{i,1},\ldots,a_{i,n_i},\ldots,a_{k,1},\ldots,a_{k,n_k}],
\end{align}
%\end{equation} 
where $n = \sum_{i=1}^s n_i$ is the total number of training vectors. The columns of the dictionary are also known as {\sl atoms}. 

We assume that there is a test vector $y \in \mathbb{R}^m$ belonging to the $i$-th class. According to~\cite{WrightYangGaneshSastryMa09}, given sufficient training samples from class $i$, $y$ can be approximately represented as a linear combination of the training samples in $A_{i}$:
\begin{equation}
y=\alpha_{i,1} x_{i,1} + \alpha_{i,2} x_{i,2} + \ldots + \alpha_{i,n_{i}} x_{i,n_{i}},
\end{equation}
where $\alpha_{i,j} \in \mathbb{R}$ are coefficients. This means if the test vector $y$ belongs to class $i$, then ideally $y$ is dependent on a small subset of training vectors $\{ a_{i,1}, \ldots, a_{i,n_i} \}$ in class $i$ only and is independent of the training vectors from all other classes. We can check whether this holds by solving the linear equation 
\begin{equation}
y = A x,  
\label{eqn:classification_equation}
\end{equation}
with unknown vector $x \in \mathbb{R}^n$ where the number of unknowns $n$ in $x$ is equal to the number of columns in the dictionary. If the ideal condition holds, $x$ has the form 
\begin{equation}
x_{\rm ideal} = [0,\ldots,0,x_{i,1},\ldots,x_{i,n_i},0,\ldots,0]^T,
\end{equation}
where $T$ denotes matrix transpose. The ideal solution $x_{\rm ideal}$ means that $y$ is a linear combination of the training vectors in $i$-th class but not others. If the ideal condition holds, $x_{\rm ideal}$ is a {\sl sparse} vector because most of its elements are zero.

\textbf{Step 2: Sparse Solution via $\ell_1$ Optimization}. According to linear algebra theory, the solution of Equation~\ref{eqn:classification_equation} depends on the following condition: if $m>n$, the system $y=Ax$ is overdetermined, and the solution can be found uniquely. However, in most applications, the number of elements in the overcomplete dictionary $A$ is typically much larger than the dimensionality of raw data (i.e., $m\ll n$). Therefore, the linear system of Equation~\ref{eqn:classification_equation} is underdetermined and has no unique solution. Since we are looking for a sparse representation $x$, we aim to solve the following $\ell_0$ optimization problem:
\begin{equation} \small
\hat{x} = \argmin_x \|x\|_0 \quad \text{ subject to } y= A x,
\label{eq:l0}
\end{equation}
where $\hat{x}$ is the sparse representation of $y$ under dictionary $A$ and $\|\cdot\|_0$ represents the $\ell_0$ norm, which counts the number of non-zero coefficients in $\hat{x}$. However, the problem with $\ell_0$ optimization is shown to be NP-hard~\cite{natarajan1995sparse}. Inspired by the recent information theory of Compressive Sensing (CS)~\cite{CandesRombergTao06,Donoho06}, the solution of $\ell_0$ optimization in Eq.~\ref{eq:l0} can be well approximated by the following $\ell_1$ optimization problem:
\begin{equation}
\hat{x} = \argmin_x \| x \|_1 \quad \text{ subject to } \|y - Ax\|_2 < \epsilon,
\label{eq:l1} 
\end{equation}
where $\epsilon$ is used to account for noise and the sparse assumption holds when the test vector can be represented by one of the classes in $A$. 

\begin{figure}
	\centering
	\subfigure[Car]{
		\includegraphics[height=7cm, width=4.1cm]{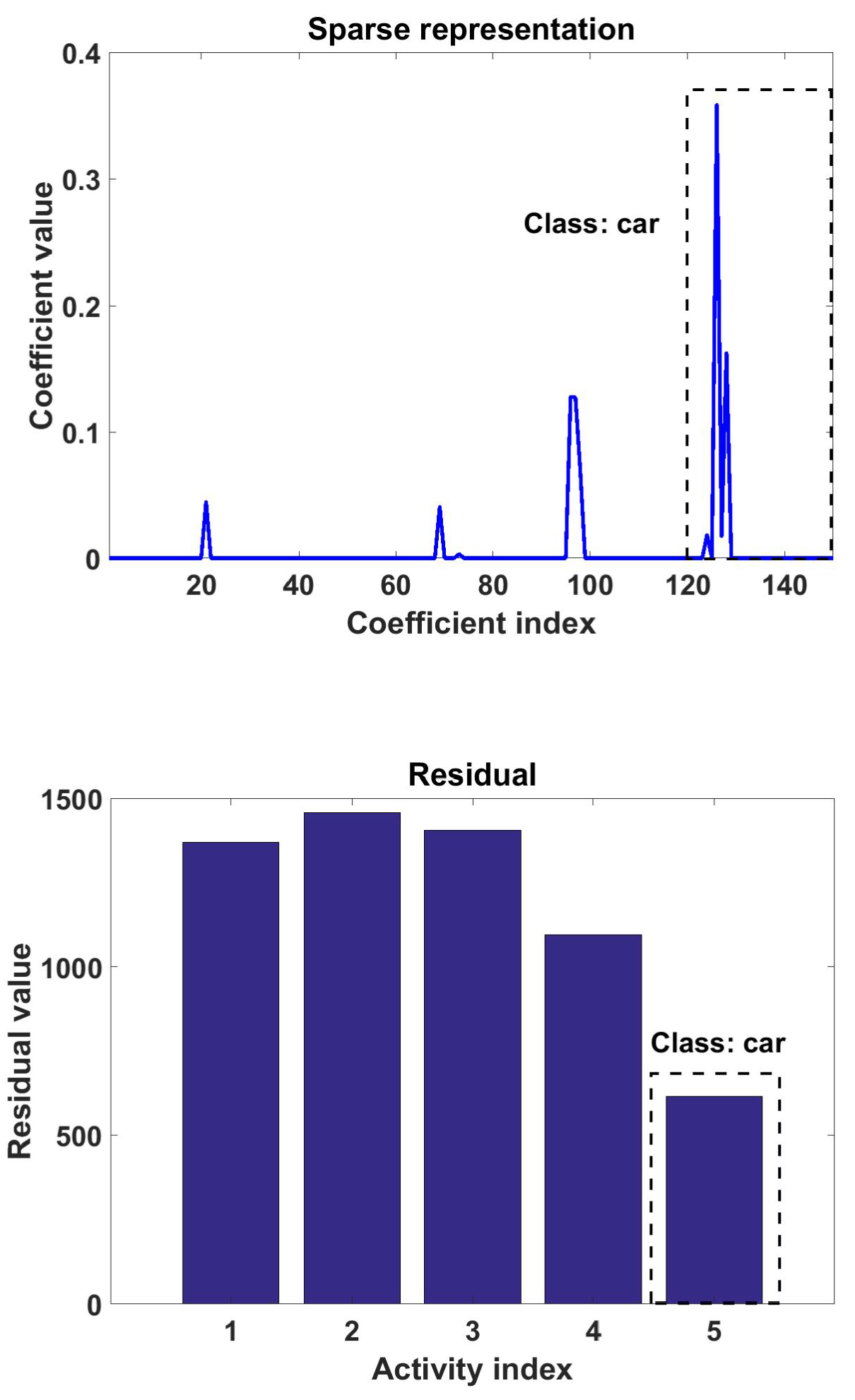}
		\label{fig:src_car}}
	\subfigure[Bus]{
		\includegraphics[height=7cm, width=4.1cm]{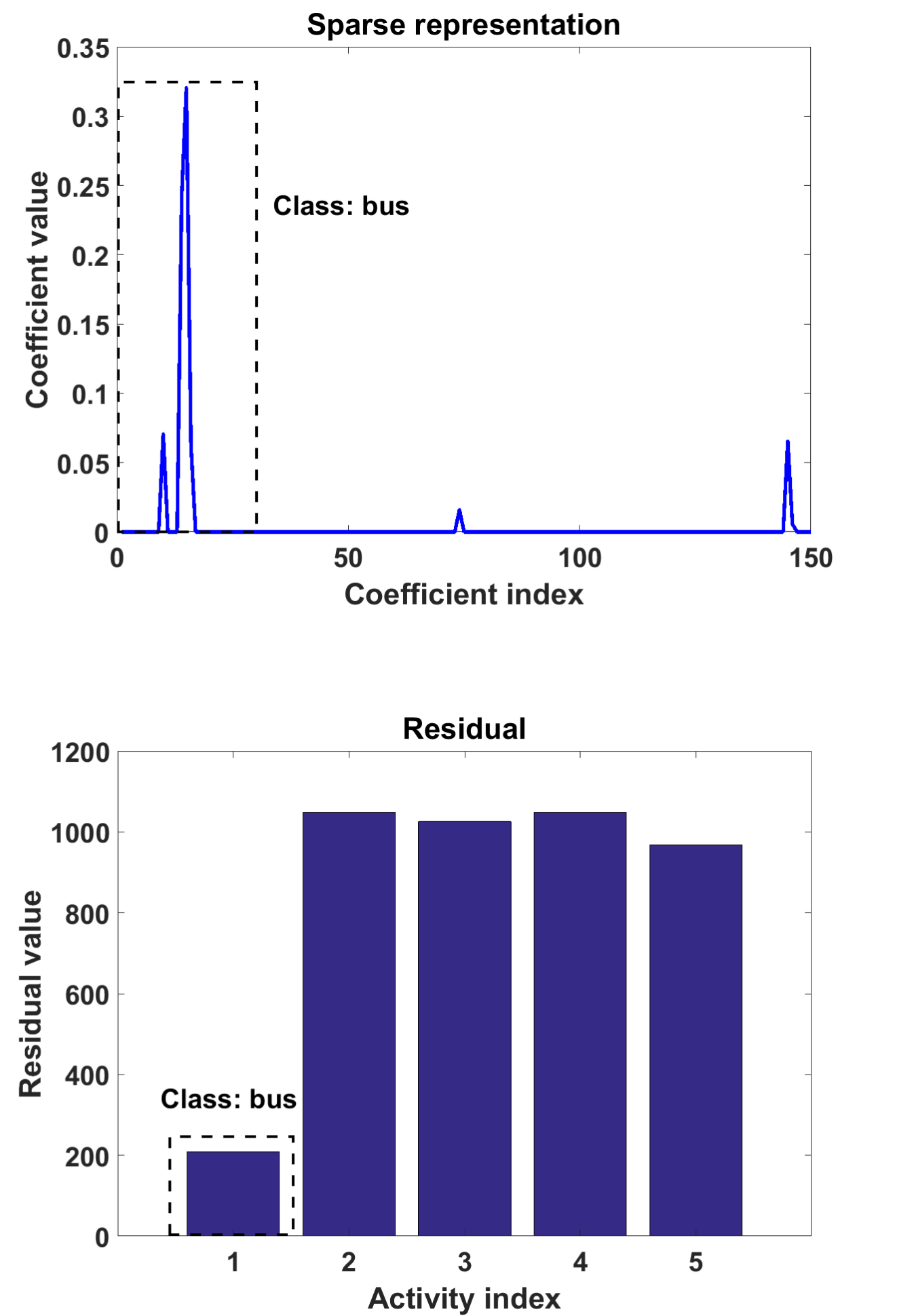}
		\label{fig:src_bus}
	}
	\caption{Sparse representation via $\ell_1$ optimization and the corresponding residuals for two test samples from car and bus, respectively.}
	\vspace{-0.1in}
\end{figure}

\textbf{Step 3: Classification}. After obtaining the sparse representation vector $\hat x\in \mathbb{R}^{n}$, the classification results can be determined by checking the residuals based on the Euclidean distance. The definition of the residual for class $i$ is:
\begin{equation}
r_i(y) = \| y - A \delta_i(\hat{x}) \|_2,
\label{eq:residual}
\end{equation}
where $\delta_i(\hat{x})\in \mathbb{R}^{N\cdot K}$ contains the coefficients related to class $i$ only (the coefficients related to other classes are set to be zeros). Then the final result of the classification will be: 
\begin{equation}
\hat{i} = \argmin_{i = 1, ...,K} r_i(y), 
\end{equation}
i.e., the right class produces the minimal residual. To illustrate this, Figure~\ref{fig:src_car} and~\ref{fig:src_bus} plots the two coefficient vectors recovered by solving Equation~\ref{eq:l1} with the noise tolerance=0.001 for two test samples from two activities: car and bus, respectively. We can observe that both of the recovered coefficient vectors are sparse. Moreover, the majority of coefficients focus on the training samples belonging to the same activity class. They also exhibit the corresponding residuals with respect to the five activity classes. We can see that both test samples are correctly classified since the minimal residual is associated with the correct activity class. 

%To conclude, the steps of the SRC can be summarized as:
%\begin{itemize}
%	\item We use K-SVD to build a dictionary $A$ which consists of the training feature vectors from different classes. 
%	\item Given a test feature vector $y$, the coefficients vector is obtained by solving Eq.~\ref{eq:l1}.
%	\item The final classification result is determined by finding the minimum compressed residual obtained from Eq.~\ref{eq:residual}. 
%\end{itemize}

%In the next section, we evaluate the proposed \SystemName.

\section{System Evaluation}
\label{evaluation}

\subsection{Hardware Platform}
%\label{sec：hardware_design}

%\begin{figure}
%	%\vspace{-0.2in}
%	\centering
%	\subfigure[Type A prototype.]{\includegraphics[scale=0.35]{figure/prototype_a.eps}}
%	\subfigure[Type B prototype.]{\includegraphics[scale=0.35]{Figure/prototype_b.eps}}\
%	\caption{Customized KEH prototypes.}
%	\label{Fig:PEHhardware}
%	\vspace{-0.1in}
%\end{figure}

\begin{figure}
	\centering
	\includegraphics[scale=0.55]{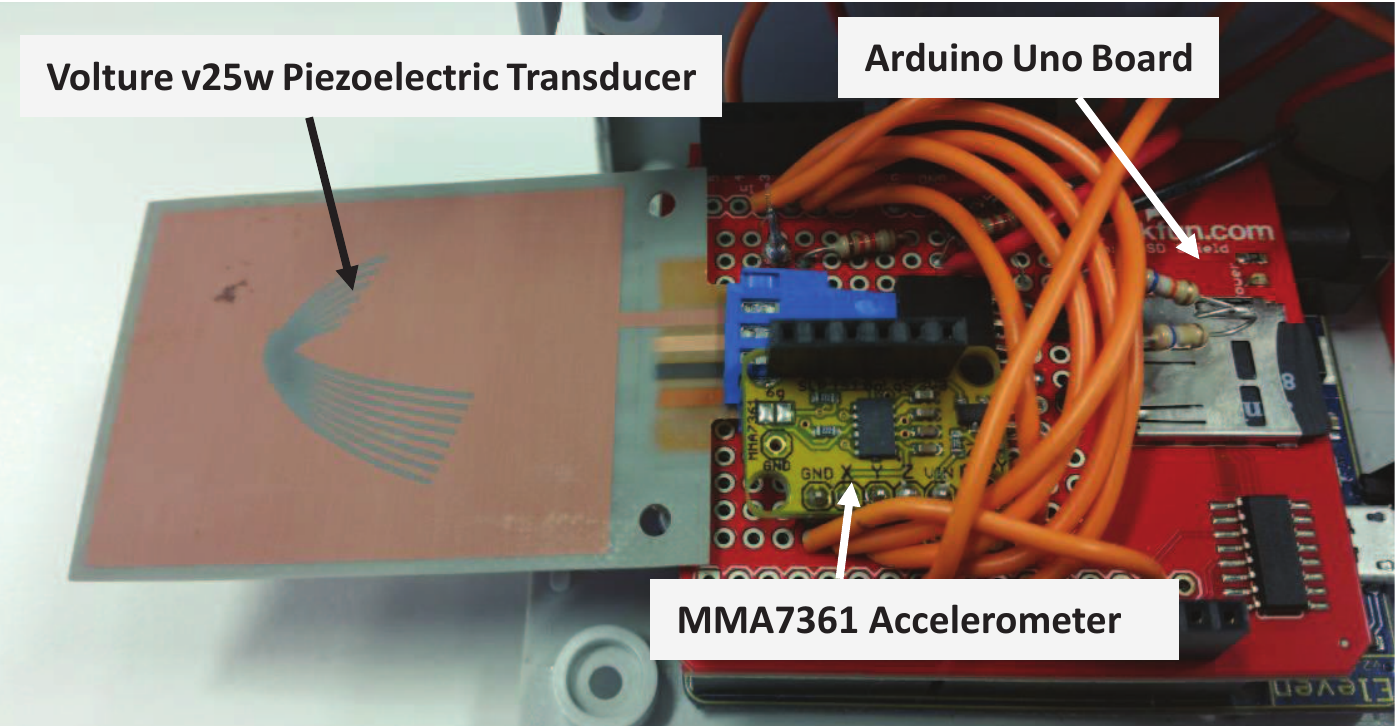}
	\caption{Customized KEH prototype.}
	\label{Fig:PEHhardware}
	\vspace{-0.1in}
\end{figure}

In this work, we build up our prototype using the off-the-shelf piezoelectric transducer from the MID\`E~\cite{MIDE}. The design of the prototype is shown in Figure~\ref{Fig:PEHhardware}, in which the V25W piezoelectric transducer is used as the KEH signal source. The form factor of the V25W transducer is 8.1cm $\times$ 3.8cm $\times$ 0.6mm. We attached it to the Arduino UNO board for data sampling purpose. The output AC voltage from the transducer is sampled by the Arduino via its onboard analog-to-digital converter (ADC) for classification purpose. Our prototypes also includes a 3-axis accelerometer (i.e., MMA7361) to measure the acceleration signals simultaneously for comparison purpose. The form-factor of our prototype is cumbersome as it is designed for the purposes of data logging and offline analysis. In a practical scenario, the form-factor of a KEH powered device can be dramatically reduced by using off-the-shelf MEMS (Micro-Electro-Mechanical Systems) energy harvesting transducers. For example, MicroGen~\cite{MicroGen} has introduced a semiconductor MEMS energy harvesting chip with a form-factor of 1.0cm$^2$ which is compatible with today's wearable and mobile devices.

%Figure~\ref{Fig:PEHhardware}(b) gives the design of the Type B prototype, differently, we utilize the MID\`E PPA-1001 model as our piezoelectric transducer. Comparing with the V25W model, PPA-1001 has a much smaller form factor and is more lightweight, which makes it able to fit within many wearable devices. In type B prototype, we applied the Texas Instrument SensorTag as the master board and sample the AC voltage generated by the PPA-1001 transducer. SensorTag is embedded with the ultra-low power ARM Cortex-M3 MCU that is widely used by today's mainstream wearable devices such as FitBit\footnote{https://www.ifixit.com/Teardown/Fitbit+Flex+Teardown/16050.}, and is running with the Contiki 3.0 operating system~\cite{ContikiOS}. A summary of those two prototypes is given in Table~\ref{table:prototype_comparison}. 

%\begin{table}
%	\centering
%	\caption{Summary of our prototypes.}
%	\label{table:prototype_comparison}
%	\small
%	\resizebox{3in}{!}{
%	\begin{tabular}{|l|c|c|}
%		\hline
%		& \textbf{Type A}& \textbf{Type B}\\ \hline
%		Master board & Arduino UNO & TI SensorTag\\
%		Piezo-transducer & MID\`E V25W & MID\`E PPA-1001\\ 
%		Resonant frequency & 160Hz & 20Hz\\
%		Signal Output & AC Voltage & AC Voltage \\
%		Accelerometer & MMA7361 & MCU9250\\ 
%		%		Sampling Rate & 100Hz & 64Hz \\
%		KEH Form factor & 8.1cm $\times$ 3.8cm & 5.5cm $\times$ 2.3cm\\\hline
%	\end{tabular}
%	}
%	%\vspace{-0.2in}
%\end{table}

\subsection{Data Collection}
\label{subsec:datacollection}

\begin{figure}
	\centering
	\includegraphics[scale=0.5]{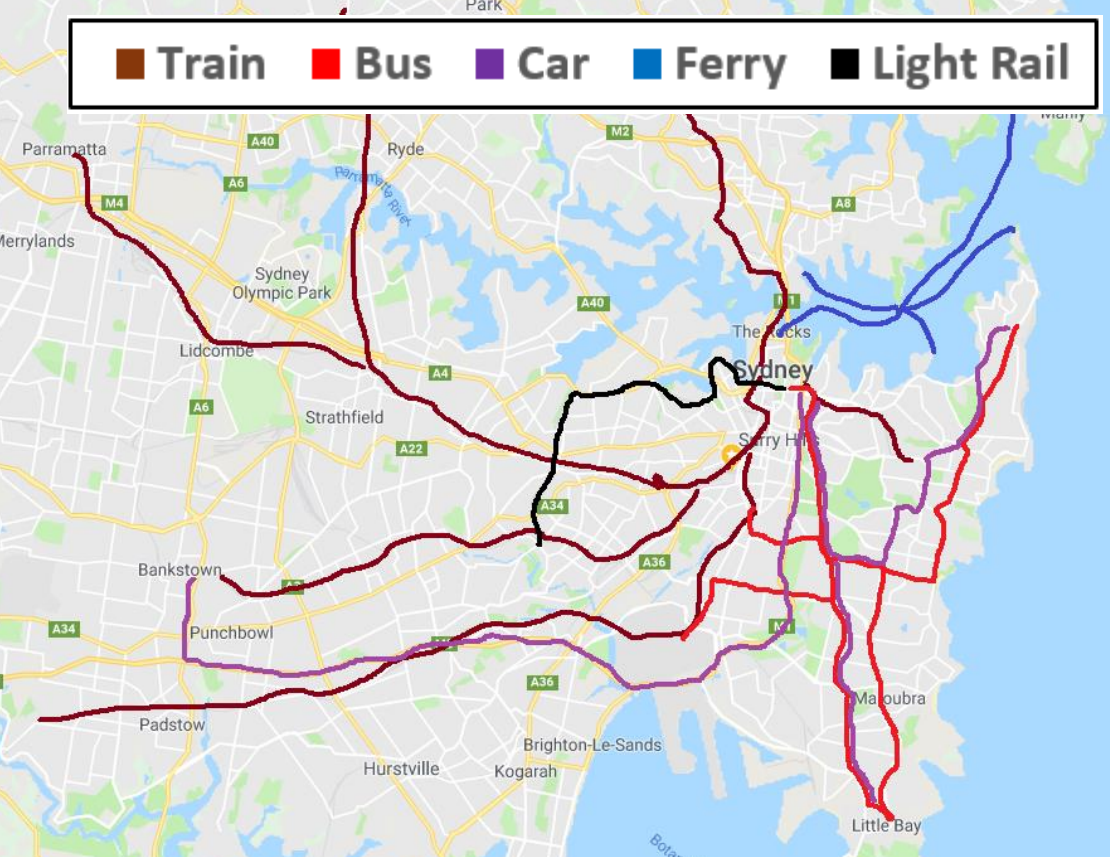}
	\caption{Examples of routes during data collection.}
	\label{Fig:routeExample}
	%\vspace{-0.1in}
\end{figure}

Our evaluation is based on a dataset collected by eight volunteers (four males and four females, height: $169.3 \pm 4.1 cm$, weight: $67.2 \pm 12.8 kg$, age: $24.6 \pm 3.7$ years) using our prototypes\footnote{Ethical approval for carrying out this experiment has been granted by the corresponding organization (Approval Number HC15888)}. Volunteers were asked to carry the prototype with them during their daily travelings. No special instructions were given about how to carry the device (i.e, position of the device), and none of the routines were decided in advance. The data collection covers a wide range of transportation modalities within our city. Figure~\ref{Fig:routeExample} indicates some of the routes the volunteers have traveled during the data collection. Table~\ref{tab:datacollection} provides a summary of the data collection. In total, over 28 hours of data have been collected which includes 97 different traces traveled by the eight volunteers during various times and traffic conditions. During data collection, both KEH and accelerometer signal are sampled at 100Hz and stored on the onboard SD card for offline analysis.

\begin{table}
	\centering
	\caption{Summary of collected data.}
	\label{tab:datacollection}
	\small
	\resizebox{2in}{!}{
		\begin{tabular}{|l|cc|}
			\hline
			& \textbf{Traces} & \textbf{Time} \\ \hline
			%Walking & 32 & 3.6 hours \\
			%Running & 32 & 2.8 hours \\
			Car & 20 & 2.4 hours \\
			Bus & 27 & 8.6 hours \\
			Train & 26 & 8.5 hours \\
			Lightrail & 16 & 4.5 hours \\ 
			Ferry & 8 & 4.2 hours \\ \hline
			\textbf{Total} & 97 & 28.2 hours \\ \hline
		\end{tabular}
	}
	\vspace{-0.1in}
\end{table}

\subsection{Goals, Metrics and Methodology}
\label{subsec:goals}
The goals of the evaluation are in two aspects: (1) Performance of \SystemName compared to accelerometer-based system; and (2) Performance of the sparse representation framework proposed in our system compared to traditional classification algorithms. We compare our motorized motion classifier with Support Vector Machine (SVM), K-Nearest Neighbor (KNN), and  Naive Bayes (NB) which are popular machine learning algorithms in activity classification. The parameters in SVM, KNN and NB are well tuned to give the highest accuracy. For each classifier, we perform 10-fold cross-validation on the collected dataset. The cross-validation process is then repeated 10 times, with each of the 10 folds used exactly once as the testing data. For fair comparison, we perform the same signal processing, feature selection and classification method on both KEH and acceleration signal. For accelerometer-based system the feature vector is obtained by concatenating features extracted along the three axes in one window together, while, for KEH-signal based approach we have only one axis AC voltage signal. In this paper, we use the true positive rate as the metric. The results are averaged with $95\%$ confidence level obtained from the 10-folds cross-validation.

\subsection{Classification Accuracy}
In the following, we examine the performance of our sparse representation based classification approach. We compare the performance of \SystemName against accelerometer-based method.

\begin{figure}[]
	\centering
	\subfigure[Accuracy given different sample rates.]{\includegraphics[scale=0.5]{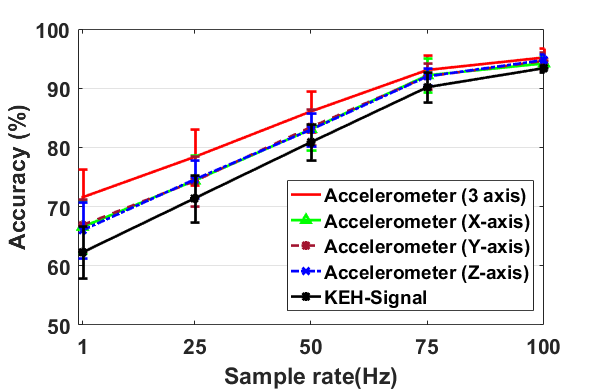}}\
	\subfigure[Accuracy given different window sizes.]{\includegraphics[scale=0.5]{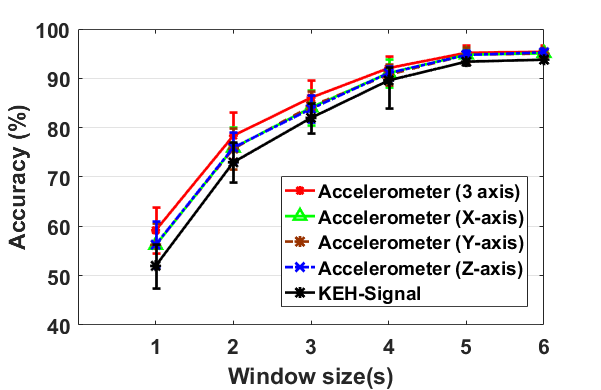}}
	\caption{Accuracy achieved by KEH and Accelerometer-based methods: (a) accuracy under different sample rates (with window size = 5s); (b) accuracy under different window sizes. The sampling rate is fixed as 75Hz and 100Hz, for accelerometer and KEH based method, respectively.} 
	\label{Fig:samplerate}
	\vspace{-0.1in}
\end{figure}

\subsubsection{\textbf{Recognition Accuracy v.s. Sampling Rate}}

The first thing we are interested in is the impact of sampling rate on the recognition accuracy, as the system power consumption is directly related to the sampling rate. We downsample the original collected KEH and acceleration data from 100Hz to several lower sampling rates (i.e., 75Hz, 50Hz, 25Hz, and 1Hz) to investigate the impact of sampling frequency.

Figure~\ref{Fig:samplerate}(a) exhibits the system accuracy of \SystemName with different sampling rates. As shown, the accuracy increases with the sampling rate for both KEH and Accelerometer-based systems (i.e., both 3-axis and single-axis). However, for accelerometer-based systems, they are able to achieve over 90\% of accuracy after the sampling rate increasing to 75Hz. On the other hand, for KEH-based method, it requires 100Hz sampling rate to achieve a comparable accuracy. Basically, this is because Accelerometer-based system can take advantage and capture more useful information for 3-axis, whereas, KEH-based system suffers from information loss due to its single axis characteristic. In addition, comparing with single-axis acceleration signal, KEH-based method still exhibits a 2\% decrease in accuracy. This is reasonable as the signal generated from the piezoelectric transducer is more noisy comparing with the single-axis acceleration signal. However, as we will discuss in Section~\ref{section:energy_consumption}, to generate low noise acceleration signal, the power consumption of Accelerometer-based system is significantly higher than that of the KEH-based system. Based on our measurement results, in terms of sensing-rated power consumption, the accelerometer-based system consumes 425$\mu$W with 75Hz sampling rate of the accelerometer in our prototype. On the other hand, to achieve similar accuracy, KEH-based system consumes 27$\mu$W with 100Hz sampling rate. This means that, in order to achieve the same system accuracy, accelerometer-based system consumes 398$\mu$W more power than the KEH-based system. In the following evaluation, we use the 100Hz and 75Hz sampling rate for KEH and accelerometer-based system, respectively.

%Lastly, Figure~\ref{Fig:samplerate}(b) exhibits the system accuracy as a function of the window size. We can notice that with a 5 seconds classification window, \SystemName can achieve over 90\% of accuracy. 

\subsubsection{\textbf{KEH v.s. Accelerometer}} 
\begin{figure}[]
	\centering
	\includegraphics[scale=0.63]{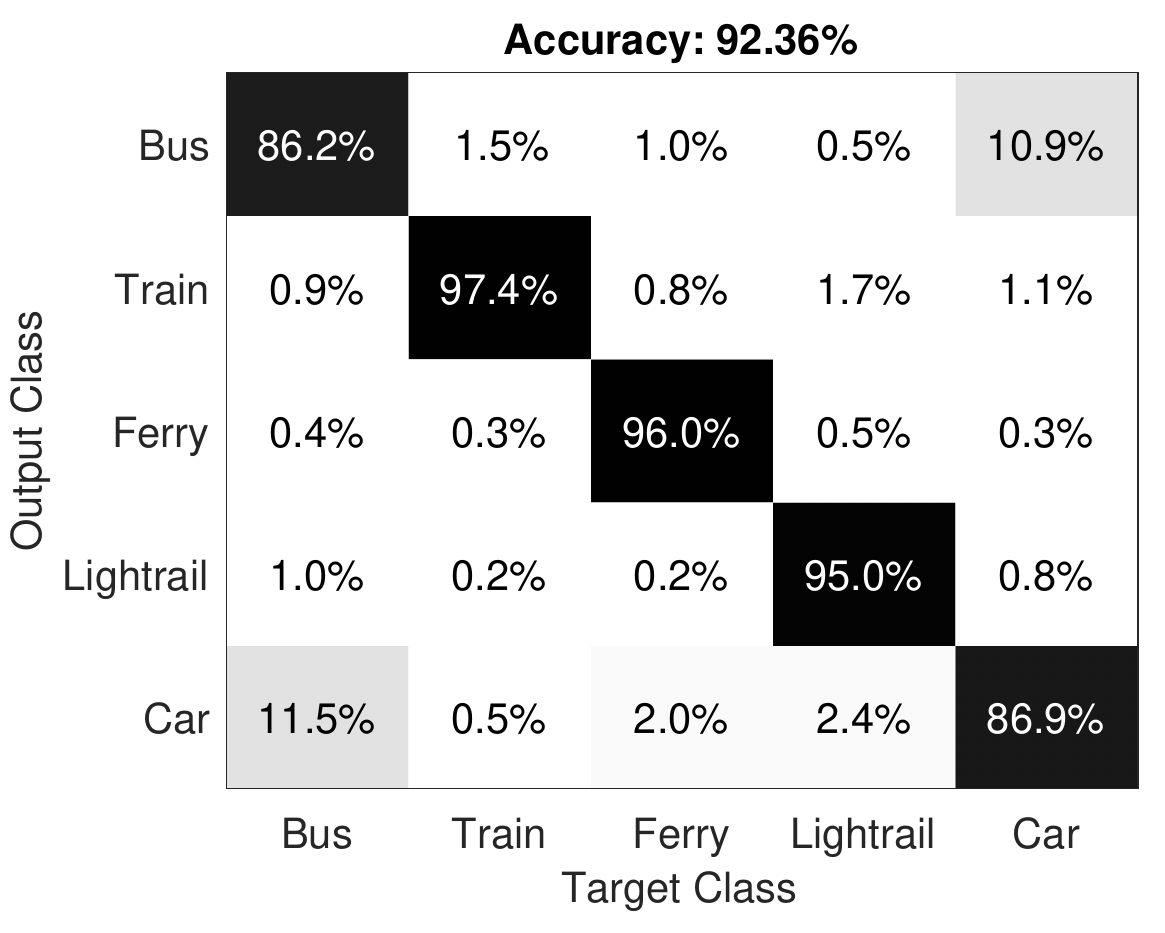}
	\caption{Confusion matrix for motorized motion classifier using KEH signal. (sampling rate is 100Hz with 5s window)}
	\label{fig:confusion}
	\vspace{-0.15in}
\end{figure}

Now, we compare the performance of \SystemName against the conventional accelerometer-based system. We vary the window size $T$ from $1s$ to $6s$ and plot the classification results in Figure~\ref{Fig:samplerate}(b). We can observe that the accuracy gap between the KEH-based method and accelerometer-based method diminishes when $T$ increases. The result is intuitive as more information can be obtained to identify the transportation modality by using more samples in a longer time window. We also find that the KEH-based method can achieve a comparable accuracy to the accelerometer-based method with $T\geq 5s$. To better characterize the performance of KEH-based method, Figure~\ref{fig:confusion} exhibits the confusion matrix of \SystemName with $T=5s$. The average accuracy achieved is 92.3\%. Moreover, we can observe that the main source of error is in the classification between the modality of bus and car. Intuitively, this difficulty is caused by the high similarity between those two transportation modalities.

%Lastly, Table~\ref{tab:end_to_end} compares the end-to-end recognition performance of KEH-based and Accelerometer-based system. As \SystemName relies on the hierarchical classification structure, the recognition results of the later stage classifiers (i.e., the pedestrian motion classifier and motorized motion classifier) are conditional to the accuracy of the earlier stage (i.e., the kinematic motion classifier). We can observe that, in case of MMC, the accuracy of KEH-based method is about 4\% lower than that of the Accelerometer-based method.  

%\begin{figure}[t]
%	\centering
%	%\includegraphics[width=2.5in, height=2in]{figure/comparisonwithaccelerometer.eps}
%	%\includegraphics[width=2.5in, height=2in]{figure/sample_rate.eps}
%	\includegraphics[height=5.6cm, width=7.5cm]{figure/sample_rate.png}
%	\caption{Accuracy with different window sizes. The sampling rate is 75Hz and 100Hz, for accelerometer and KEH based method, respectively.} 
%	\label{fig:comp_acc}
%	%\vspace{-0.1in}
%\end{figure}

%\begin{table}[]
%	\centering
%	\caption{Confusion matrix for motorized motion classifier using KEH signal. (sampling rate is 100Hz with 5s window)}
%	\label{tab:confusion}
%	\small
%	\resizebox{3.4in}{!}{
%		\begin{tabular}{|c|c|c|c|c|c|}
%			\hline
%			& \textbf{Bus}    & \textbf{Train}    & \textbf{Ferry}  & \textbf{Lightrail} & \textbf{Car}  \\ \hline
%			\textbf{Bus}       & 86.5\% & 1.5\%  & 1\%    & 0.5\%     & 10.5\%  \\ \hline
%			\textbf{Train}       & 0.9\%  & 95.5\% & 0.85\% & 1.7\%     & 1.05\% \\ \hline
%			\textbf{Ferry}     & 0.4\%  & 0.34\% & 98.5\% & 0.5\%     & 0.26\% \\ \hline
%			\textbf{Lightrail} & 1\%    & 0.23   & 0.21\% & 97.8\%    & 0.76\% \\ \hline
%			\textbf{Car}     & 11.5\% & 0.5\%  & 2\%    & 2.5\%     & 83.5\% \\ \hline
%		\end{tabular}
%	}
%	%\vspace{-0.1in}
%\end{table}

%\begin{table}[]
%	\centering
%	\caption{Summary of end-to-end recognition accuracy.}
%	%\small
%	\label{tab:end_to_end}
%	\resizebox{3.4in}{!}{
%		\begin{tabular}{|l|cc|}
%			\hline
%			Classifier & \textbf{KEH} & \textbf{Accelerometer} \\ \hline \hline
%			KMC & 95.32 & 97.50  \\ \hline \hline
%			PMC & 95.32$\times$98.22 = 93.62 & 97.50$\times$98.85 = 96.38 \\ 
%			MMC & 95.32$\times$92.30 = 87.89 & 97.50$\times$94.30 = 91.94 \\ \hline
%		\end{tabular}
%	}
%	%\vspace{-0.1in}
%\end{table}

\subsubsection{\textbf{Comparison with other classifiers}} 
\begin{figure}[]
	\centering
	\includegraphics[scale=0.28]{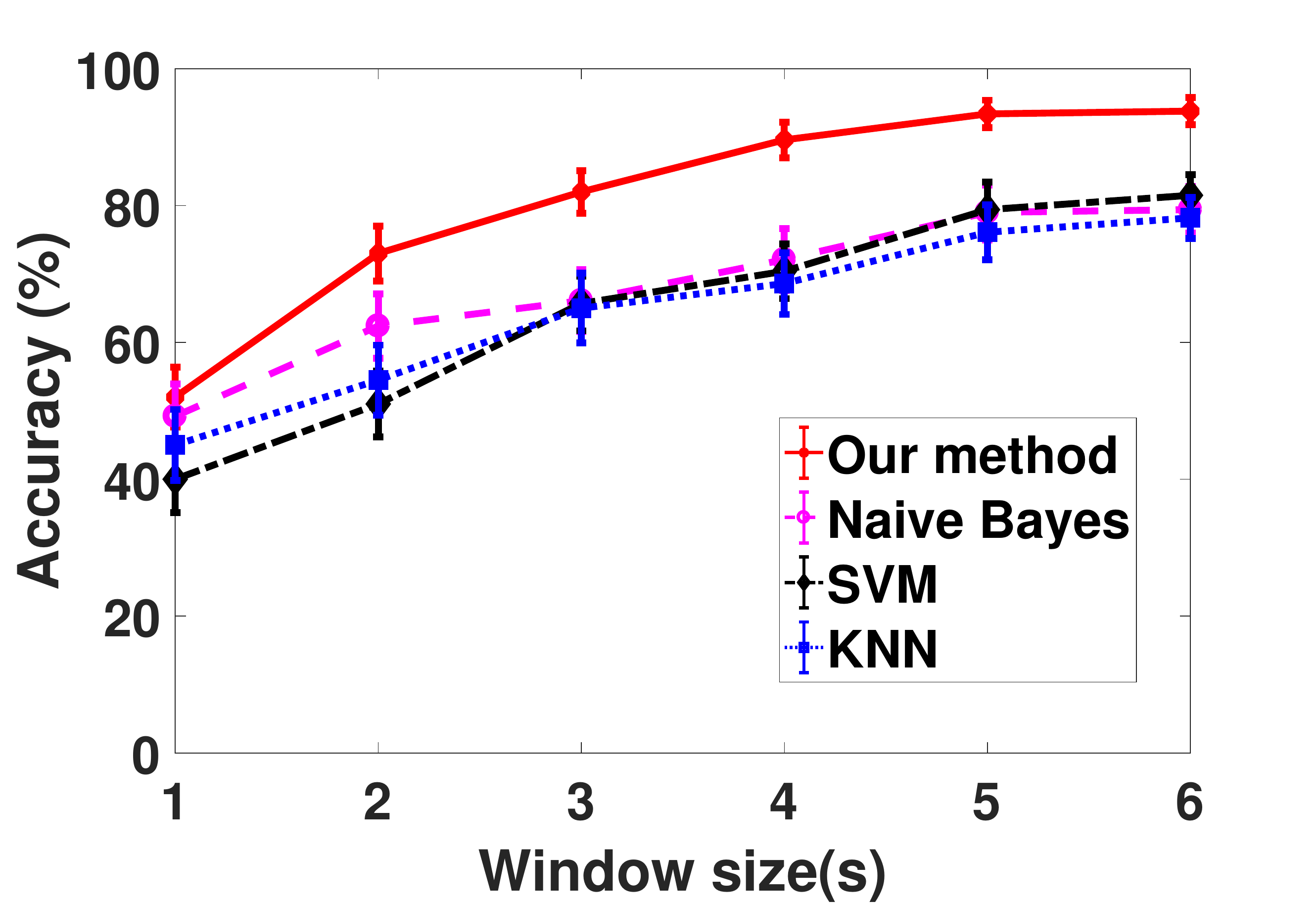}
	\caption{Accuracy of KEH-based system with different classifiers. (with 100Hz sampling rate)} 
	\label{fig:comp_others}
	\vspace{-0.1in}
\end{figure}

In the following, we compare the proposed sparse representation-based classifier against traditional classifiers. We use the same features and vary the window time $T$ from $1s$ to $6s$. As shown in Figure~\ref{fig:comp_others}, we can observe that our approach is up to $10\%$ more accurate than the best traditional classifier (i.e., SVM with $T\geq 5s$). This accuracy improvement is coming from the use of the dictionary learning based method as it provides more compact representation of the activities while preserving richer information, thereby underpinning higher recognition performance in transportation mode detection. 

\subsubsection{\textbf{Comparison with Baseline System}} 
Lastly, we compare our results with the baseline system proposed by Hemminiki et al.~\cite{hemminki2013accelerometer}. In the work of Hemminiki et al., they applied the similar features that we used in this paper. However, the overall accuracy they can achieve is approximately 85\%, which is approximately 10\% lower than the performance of \SystemName. There are two reasons for this difference. First, Hemminiki et al. applied only conventional classifiers in their system. As shown in Figure~\ref{fig:comp_others}, comparing with our dictionary-based sparse representation classifier, traditional classifiers suffer at least 10\% accuracy loss. Secondly, Hemminiki et al. did not filter the signal generated during the vehicle stop/pause periods out from the acceleration signal. As we have shown, the motorized motions naturally include some stationary periods within which the vehicle is stopped. Consequently, it generates more errors in distinguishing the `Stationary' class from these motorized classes in the presence of the stationary periods.%}  

\subsection{System Robustness}
A practical challenge in transportation mode detection is the robustness of the system against different kinds of variation. In the following, we consider two major variations: the variance results from different traveling traces and the variance due to user difference.

\subsubsection{\textbf{Robustness to Trace Variance}}
The trace variance may come from the user traveling on a different routine, traveling by a different vehicle, or with different traffic conditions, and many others. To evaluate our system robustness against trace variance, we carry out \textbf{\textit{leave-one-trace-out}} cross validation over the data traces we have collected. The averaged results are shown in Table~\ref{tab:trace}. As expected, for both KEH and Accelerometer-based systems, the recognition accuracy decreases compared to the results reported in Figure~\ref{Fig:samplerate}. However, the accuracy only slightly drops by 2.7\% for KEH-based system which demonstrates its robustness to the trace variance.

\begin{table}
	\centering
	\caption{Results of the robustness experiments.}
	\label{tab:trace}
	\small
	\resizebox{3in}{!}{
		\begin{tabular}{|l|cc|}
			\hline
			 & \textbf{KEH} & \textbf{Accelerometer} \\ \hline
			\textbf{Trace Variance} & 89.6$\pm$0.84 (-2.7) & 92.5$\pm$0.67 (-1.8) \\
			\textbf{User Variance} & 85.1$\pm$0.74 (-7.2) & 89.5$\pm$0.67 (-6.4) \\ \hline
		\end{tabular}
	}
	\vspace{-0.1in}
\end{table}

\subsubsection{\textbf{Robustness to User Variance}}

Similar, to evaluate the robustness of \SystemName against user variance, we carry out \textbf{\textit{leave-one-user-out}} cross validation over the data collected by the eight volunteers. In particular, we use data from one user as the testing set and the remaining data collected from the other seven subjects as the training set. The results are averaged across the eight subjects. As indicated in Table~\ref{tab:trace}, we can notice that the user variance has a higher impact on classification accuracy than the trace variance. Both KEH and Accelerometer have suffered more than 6\% accuracy drop. This is because different users tend to carry the device in different ways during the data collection (e.g., hold in the hand or put it in the backpack). This suggests that a subject-dependent model may be considered in the practical design to achieve better performance. Overall, \SystemName achieves over 85\% of accuracy given difference variances, which is comparable to the results reported in the literature~\cite{hemminki2013accelerometer,shin2015urban}.

\subsection{Energy Consumption}
\label{section:energy_consumption}

High energy consumption is widely regarded as the major barrier in achieving long term sensing~\cite{seneviratne2017survey}. The major energy consumption of a wearable sensing system comes from two parts: \textit{data sampling}, and \textit{data transmission}. Fortunately, thanks to the current advancements in \textit{ambient backscatter communication}~\cite{liu2013ambient}, the power consumption in wireless data transmission can be potentially reduced to less than 1$\mu W$ while achieving over 1Kbps data rate~\cite{liu2013ambient}. Thus, data sampling is expected to become the lion's share of the system power consumption for future low-power wearable systems. In the following, we present a power consumption profiling of state-of-the-art low-power device to understand the power consumption in data sampling of accelerometer-based transportation mode detection system, and demonstrate the advantage of \SystemName in energy saving.

\subsubsection{Measurement Setup}

%\begin{figure}[]
%	\centering
%	\includegraphics[height=5cm, width=7.3cm]{figure/measurement_setup_raw.pdf}
%	\caption{Power measurement setup. The accelerometers (i.e., ADXL335, ADXL362, and ADXL377) are connected to SensorTag ADC/SPI via external wires.} 
%	\label{Fig:measurement_setup}
%	%\vspace{-0.1in}
%\end{figure}

We select the Texas Instrument CC2650 SensorTag as the target device, which is an ideal representative of current low-power IoT devices. We investigate the power consumption of SensorTag in sampling the voltage signal from the KEH transducer, and that in sampling acceleration signal from accelerometers. We consider four different models of low-power accelerometer in the market: MPU9250, ADXL335, ADXL377, and ADXL362. The MPU9250 is originally embedded with the SensorTag. It is a 9-axis digital sensor, combining gyroscope, digital compass, and accelerometer. During the power measurements, we only enable the 3-axis accelerometer and leave all the other two sensors turned off. The signal of MPU9250 is sampled using the Inter-Integrated Circuit (I$^2$C) bus. For the two analog accelerometers, ADXL335 and ADXL377, we sampled their signal through an external wire connected to the ADC on the SensorTag. Lastly, the digital ADXL362 is sampled through the Serial Peripheral Interface (SPI) bus on SensorTag. The SensorTag is running with the Contiki OS, in which the MCU is duty-cycled to save power. The Agilent DSO3202A oscilloscope is used to measure the power consumption. %The setup of the power measurement is given in Figure~\ref{Fig:measurement_setup}. 

%Moreover, all unnecessary components, including the onboard ADC, SPI bus, and the accelerometers are powered-off when it is not sampling. Expect that ADXL362 is always-on in the experiment. Given its ultra-low power consumption in the motion activated wake-up mode (additional 0.9$\mu$W to the baseline system power consumption.), it would not be necessary to turn it off~\cite{qi2016gazelle}. 

\subsubsection{Power Consumption Analysis}

\begin{table}[t]
	\small
	\centering
	\caption{Power consumption ($\mu$W) in data sampling given different sampling frequencies (Hz).}
	\label{table:power_consumption}
	\resizebox{3.3in}{!}{
		\begin{tabular}{|c|c|c|c|c|c|}
			\hline
			& \textbf{KEH}& \textbf{ADXL335}& \textbf{ADXL377}& \textbf{MPU9250} & \textbf{ADXL362}\\ \hline
			1Hz & 6 & 7 & 8 & 11 & 6\\
			25Hz & 11 & 39 & 51 & 145 & 18\\
			50Hz & 16 & 72 & 96 & 285 & 29\\
			75Hz & 21 & 105 & 141 & 425 & \textbf{41}\\
			100Hz &\textbf{27} & 138 & 186 & 633 & 52\\\hline
		\end{tabular}}
		\vspace{-0.1in}
\end{table}
	
Table~\ref{table:power_consumption} gives the power consumption in data sampling with different sampling frequencies. As shown, \SystemName achieves significant power saving in data sampling, comparing with the conventional accelerometer-based system. More specifically, recall the results shown in Figure~\ref{Fig:samplerate}, for KEH-based system, a sampling rate of 100Hz is required to achieve high recognition accuracy. Therefore, \SystemName consumes 27$\mu$W in data sampling. On the other hand, for accelerometer-based system, to achieve the similar accuracy, only 75Hz sampling rate is needed. We can observe that the lowest power consumption of accelerometer-based system is 41$\mu$W at 75Hz in terms of ADXL362 (which is widely regarded as the most power-efficient accelerometer in current industry), while it consumes as high as 425$\mu$W for MPU9250. This means that, given different accelerometers, to achieve the same classification accuracy, the accelerometer-based system needs to consume 52\% (i.e., ADXL362) to 1474\% (i.e., MPU9250) more power than \SystemName in data sampling.

\section{Conclusion and Future Directions}
\label{sec:conclusion}
	
In this paper, we present \SystemName which leverages KEH voltage signal for energy-efficient transportation mode sensing. A sparse representation based framework is designed to improve the system accuracy. We evaluate the proposed system using over 28 hours of transportation data collected by a customized KEH prototype. Evaluation results indicate that our sparse representation-based approach achieves $92\%$ of accuracy which is about $10\%$ better than traditional classification algorithms such as kNN and SVM. In addition, comparing to accelerometer-based system, EnTrans is 34\% more power efficient without sacrificing classification performance.

While our work clearly demonstrates the usefulness of EnTrans in classifying different daily transportation modes, there are several future directions to investigate its full potential. First, different from conventional motion sensors that provide three-axises of signal, the KEH transducer used in our prototype can only provide single axis signal. As a future direction, we can take advantage of the recently proposed multi-axis energy harvesters~\cite{aktakka2015three,hung2015miniature} to acquire much richer information for sensing. Second, although the use of sparse representation based classification can provide better recognition performance, its computational complexity is high. Moreover, to ensure robustness against user and traveling route variances, a large training dictionary is required. We plan to explore more advanced machine learning algorithms, such as deep learning~\cite{liang2017convolutional,fang2017learning} to resolve this limitation. For instance, we can exploit the deep convolutional neural networks (CNN)~\cite{lecun1998gradient} to learn a proper data representation (i.e., features) from the KEH signal that are more robust compared to the manually crafted ones~\cite{zhou2016learning,xiao2016learning}. In addition, with the help of transfer learning~\cite{yosinski2014transferable,xie2016transfer}, we can leverage the low-level features that learned from the dataset of a known environment (i.e., data collected from a particluar city) to augment the learning task for a new environment. Such that, we can adaptively improve and generalize our system model for diverse environments and users, and thus, ensure more stable classification performance.

%\balance
\bibliographystyle{IEEEtran}
\bibliography{sigproc}

\begin{thebibliography}{10}
\providecommand{\url}[1]{#1}
\csname url@samestyle\endcsname
\providecommand{\newblock}{\relax}
\providecommand{\bibinfo}[2]{#2}
\providecommand{\BIBentrySTDinterwordspacing}{\spaceskip=0pt\relax}
\providecommand{\BIBentryALTinterwordstretchfactor}{4}
\providecommand{\BIBentryALTinterwordspacing}{\spaceskip=\fontdimen2\font plus
\BIBentryALTinterwordstretchfactor\fontdimen3\font minus
  \fontdimen4\font\relax}
\providecommand{\BIBforeignlanguage}[2]{{%
\expandafter\ifx\csname l@#1\endcsname\relax
\typeout{** WARNING: IEEEtran.bst: No hyphenation pattern has been}%
\typeout{** loaded for the language `#1'. Using the pattern for}%
\typeout{** the default language instead.}%
\else
\language=\csname l@#1\endcsname
\fi
#2}}
\providecommand{\BIBdecl}{\relax}
\BIBdecl

\bibitem{seneviratne2017survey}
S.~Seneviratne, Y.~Hu, T.~Nguyen, G.~Lan, S.~Khalifa, K.~Thilakarathna,
  M.~Hassan, and A.~Seneviratne, ``A survey of wearable devices and
  challenges,'' \emph{IEEE Communications Surveys \& Tutorials}, vol.~19,
  no.~4, pp. 2573--2620, 2017.

\bibitem{zheng2008understanding}
Y.~Zheng, Q.~Li, Y.~Chen, X.~Xie, and W.-Y. Ma, ``Understanding mobility based
  on {GPS} data,'' in \emph{Proceedings of the ACM International Conference on
  Ubiquitous computing}, 2008.

\bibitem{hemminki2013accelerometer}
S.~Hemminki, P.~Nurmi, and S.~Tarkoma, ``Accelerometer-based transportation
  mode detection on smartphones,'' in \emph{Proceedings of the ACM Conference
  on Embedded Networked Sensor Systems}, 2013.

\bibitem{froehlich2009ubigreen}
J.~Froehlich, T.~Dillahunt, P.~Klasnja, J.~Mankoff, S.~Consolvo, B.~Harrison,
  and J.~A. Landay, ``Ubigreen: Investigating a mobile tool for tracking and
  supporting green transportation habits,'' in \emph{Proceedings of the
  {SIGCHI} Conference on Human Factors in Computing Systems}, 2009.

\bibitem{mitcheson2008energy}
P.~D. Mitcheson, E.~M. Yeatman, G.~K. Rao, A.~S. Holmes, and T.~C. Green,
  ``Energy harvesting from human and machine motion for wireless electronic
  devices,'' \emph{Proceedings of the IEEE}, vol.~96, no.~9, pp. 1457--1486,
  2008.

\bibitem{huang2015battery}
Q.~Huang, Y.~Mei, W.~Wang, and Q.~Zhang, ``Battery-free sensing platform for
  wearable devices: The synergy between two feet,'' in \emph{Proceedings of the
  Annual IEEE International Conference on Computer Communications}, 2015.

\bibitem{AMPY}
\BIBentryALTinterwordspacing
AMPY. [Online]. Available: \url{{http://www.getampy.com/ampy-move.html/},}
\BIBentrySTDinterwordspacing

\bibitem{SEQUENT}
\BIBentryALTinterwordspacing
Sequent. [Online]. Available: \url{http://www.sequentwatch.com/}
\BIBentrySTDinterwordspacing

\bibitem{MicroGen}
\BIBentryALTinterwordspacing
Microgen {BLOT} energy harvester. [Online]. Available:
  \url{https://www.prweb.com/releases/2011/6/prweb8585499.htm}
\BIBentrySTDinterwordspacing

\bibitem{rao2014systems}
V.~Rao, S.~A.~U. Nambi, R.~Prasad, and I.~Niemegeers, ``On systems generating
  context triggers through energy harvesting,'' \emph{IEEE Communications
  Magazine}, vol.~52, no.~6, pp. 70--77, 2014.

\bibitem{khalifa2017harke}
S.~Khalifa, G.~Lan, M.~Hassan, A.~Seneviratne, and S.~K. Das, ``{HARKE}: Human
  activity recognition from kinetic energy harvesting data in wearable
  devices,'' \emph{IEEE Transactions on Mobile Computing}, vol.~17, no.~6, pp.
  1353--1368, 2018.

\bibitem{xiang2013powering}
T.~Xiang, Z.~Chi, F.~Li, J.~Luo, L.~Tang, L.~Zhao, and Y.~Yang, ``Powering
  indoor sensing with airflows: A trinity of energy harvesting, synchronous
  duty-cycling, and sensing,'' in \emph{Proceedings of the ACM Conference on
  Embedded Networked Sensor Systems}, 2013.

\bibitem{campbell2016perpetual}
B.~Campbell, M.~Clark, S.~DeBruin, B.~Ghena, N.~Jackson, Y.-S. Kuo, and
  P.~Dutta, ``perpetual sensing for the built environment,'' \emph{IEEE
  Pervasive Computing}, vol.~15, no.~4, pp. 45--55, 2016.

\bibitem{mahbub2018computer}
M.~Hassan, W.~Hu, G.~Lan, A.~Seneviratne, S.~Khalifa, and S.~K. Das,
  ``Kinetic-powered health wearables: Challenges and opportunities,''
  \emph{Computer}, vol.~51, no.~9, pp. 64--74, 2018.

\bibitem{kalantarian2015monitoring}
H.~Kalantarian, N.~Alshurafa, T.~Le, and M.~Sarrafzadeh, ``Monitoring eating
  habits using a piezoelectric sensor-based necklace,'' \emph{Computers in
  Biology and Medicine}, vol.~58, pp. 46--55, 2015.

\bibitem{lan2016transportation}
G.~Lan, W.~Xu, S.~Khalifa, M.~Hassan, and W.~Hu, ``Transportation mode
  detection using kinetic energy harvesting wearables,'' in \emph{Proceedings
  of the IEEE International Conference on Pervasive Computing and Communication
  Workshops}, 2016.

\bibitem{patterson2003inferring}
D.~J. Patterson, L.~Liao, D.~Fox, and H.~Kautz, ``Inferring high-level behavior
  from low-level sensors,'' in \emph{Proceedings of the International
  Conference on Ubiquitous Computing}, 2003.

\bibitem{reddy2010using}
S.~Reddy, M.~Mun, J.~Burke, D.~Estrin, M.~Hansen, and M.~Srivastava, ``Using
  mobile phones to determine transportation modes,'' \emph{ACM Transactions on
  Sensor Networks}, vol.~6, no.~2, p.~13, 2010.

\bibitem{stenneth2011transportation}
L.~Stenneth, O.~Wolfson, P.~S. Yu, and B.~Xu, ``Transportation mode detection
  using mobile phones and gis information,'' in \emph{Proceedings of the ACM
  SIGSPATIAL International Conference on Advances in Geographic Information
  Systems}, 2011.

\bibitem{sohn2006mobility}
T.~Sohn, A.~Varshavsky, A.~LaMarca, M.~Y. Chen, T.~Choudhury, I.~Smith,
  S.~Consolvo, J.~Hightower, W.~G. Griswold, and E.~De~Lara, ``Mobility
  detection using everyday {GSM} traces,'' in \emph{Proceedings of the ACM
  International Conference on Ubiquitous Computing}, 2006.

\bibitem{muller2006practical}
I.~A.~H. Muller, ``Practical activity recognition using {GSM} data,'' in
  \emph{Proceedings of the International Semantic Web Conference}, 2006.

\bibitem{mun2008parsimonious}
M.~Mun, D.~Estrin, J.~Burke, and M.~Hansen, ``Parsimonious mobility
  classification using {GSM} and {Wi-Fi} traces,'' in \emph{Proceedings of the
  ACM Workshop on Embedded Networked Sensors}, 2008.

\bibitem{muthukrishnan2007sensing}
K.~Muthukrishnan, M.~Lijding, N.~Meratnia, and P.~Havinga, ``Sensing motion
  using spectral and spatial analysis of {WLAN RSSI},'' in \emph{Proceedings of
  the European Conference on Smart Sensing and Context}.\hskip 1em plus 0.5em
  minus 0.4em\relax Springer, 2007.

\bibitem{sankaran2014using}
K.~Sankaran, M.~Zhu, X.~F. Guo, A.~L. Ananda, M.~C. Chan, and L.-S. Peh,
  ``Using mobile phone barometer for low-power transportation context
  detection,'' in \emph{Proceedings of the ACM Conference on Embedded Networked
  Sensor Systems}, 2014.

\bibitem{yu2014big}
M.-C. Yu, T.~Yu, S.-C. Wang, C.-J. Lin, and E.~Y. Chang, ``Big data small
  footprint: the design of a low-power classifier for detecting transportation
  modes,'' \emph{Proceedings of the VLDB Endowment}, vol.~7, no.~13, pp.
  1429--1440, 2014.

\bibitem{liang2017convolutional}
X.~Liang and G.~Wang, ``A convolutional neural network for transportation mode
  detection based on smartphone platform,'' in \emph{Proceedings of the IEEE
  International Conference on Mobile Ad Hoc and Sensor Systems}, 2017.

\bibitem{fang2017learning}
S.-H. Fang, Y.-X. Fei, Z.~Xu, and Y.~Tsao, ``Learning transportation modes from
  smartphone sensors based on deep neural network,'' \emph{IEEE Sensors
  Journal}, vol.~17, no.~18, pp. 6111--6118, 2017.

\bibitem{lan2015estimating}
G.~Lan, S.~Khalifa, M.~Hassan, and W.~Hu, ``Estimating calorie expenditure from
  output voltage of piezoelectric energy harvester-an experimental feasibility
  study,'' in \emph{Proceedings of the EAI International Conference on Body
  Area Networks}, 2015.

\bibitem{weitao2016ndss}
W.~Xu, G.~Lan, Q.~Lin, S.~Khalifa, N.~Bergmann, M.~Hassan, and H.~Wen,
  ``{KEH-Gait}: Towards a mobile healthcare user authentication system by
  kinetic energy harvesting,'' in \emph{Proceedings of the Network and
  Distributed System Security Symposium}, 2017.

\bibitem{blank2016ball}
P.~Blank, T.~Kautz, and B.~M. Eskofier, ``Ball impact localization on table
  tennis rackets using piezo-electric sensors,'' in \emph{Proceedings of the
  ACM International Symposium on Wearable Computers}, 2016.

\bibitem{ma2018sehs}
D.~Ma, G.~Lan, W.~Xu, M.~Hassan, and W.~Hu, ``{SEHS}: Simultaneous energy
  harvesting and sensing using piezoelectric energy harvester,'' in
  \emph{Proceedings of the IEEE/ACM International Conference on
  Internet-of-Things Design and Implementation}, 2018.

\bibitem{lan2017veh}
G.~Lan, W.~Xu, S.~Khalifa, M.~Hassan, and W.~Hu, ``{VEH-COM}: Demodulating
  vibration energy harvesting for short range communication,'' in
  \emph{Proceedings of the IEEE International Conference on Pervasive Computing
  and Communications}, 2017.

\bibitem{lan2018hidden}
G.~Lan, D.~Ma, M.~Hassan, and W.~Hu, ``{HiddenCode}: Hidden acoustic signal
  capture with vibration energy harvesting,'' in \emph{Proceedings of the IEEE
  International Conference on Pervasive Computing and Communications}, 2018.

\bibitem{bhatti2016energy}
N.~A. Bhatti, M.~H. Alizai, A.~A. Syed, and L.~Mottola, ``Energy harvesting and
  wireless transfer in sensor network applications: Concepts and experiences,''
  \emph{ACM Transactions on Sensor Networks}, vol.~12, no.~3, p.~24, 2016.

\bibitem{shi2016edge}
W.~Shi, J.~Cao, Q.~Zhang, Y.~Li, and L.~Xu, ``Edge computing: Vision and
  challenges,'' \emph{IEEE Internet of Things Journal}, vol.~3, no.~5, pp.
  637--646, 2016.

\bibitem{datta2016integrating}
S.~K. Datta, R.~P.~F. Da~Costa, J.~H{\"a}rri, and C.~Bonnet, ``Integrating
  connected vehicles in internet of things ecosystems: Challenges and
  solutions,'' in \emph{Proceedings of the IEEE International Symposium on A
  World of Wireless, Mobile and Multimedia Networks}, 2016.

\bibitem{stockx2014subwayps}
T.~Stockx, B.~Hecht, and J.~Sch{\"o}ning, ``Subwayps: Towards smartphone
  positioning in underground public transportation systems,'' in
  \emph{Proceedings of the ACM SIGSPATIAL International Conference on Advances
  in Geographic Information Systems}, 2014.

\bibitem{thiagarajan2010cooperative}
A.~Thiagarajan, J.~Biagioni, T.~Gerlich, and J.~Eriksson, ``Cooperative transit
  tracking using smart-phones,'' in \emph{Proceedings of the ACM Conference on
  Embedded Networked Sensor Systems}, 2010.

\bibitem{bulling2014tutorial}
A.~Bulling, U.~Blanke, and B.~Schiele, ``A tutorial on human activity
  recognition using body-worn inertial sensors,'' \emph{ACM Computing Surveys},
  vol.~46, no.~3, p.~33, 2014.

\bibitem{liu1998feature}
H.~Liu and H.~Motoda, \emph{Feature extraction, construction and selection: A
  data mining perspective}.\hskip 1em plus 0.5em minus 0.4em\relax Springer
  Science \& Business Media, 1998, vol. 453.

\bibitem{schutze2008introduction}
H.~Sch{\"u}tze, C.~D. Manning, and P.~Raghavan, \emph{Introduction to
  information retrieval}.\hskip 1em plus 0.5em minus 0.4em\relax Cambridge
  University Press, 2008, vol.~39.

\bibitem{frank2013touchalytics}
M.~Frank, R.~Biedert, E.~Ma, I.~Martinovic, and D.~Song, ``Touchalytics: On the
  applicability of touchscreen input as a behavioral biometric for continuous
  authentication,'' \emph{IEEE Transactions on Information Forensics and
  Security}, vol.~8, no.~1, pp. 136--148, 2013.

\bibitem{ding2005minimum}
C.~Ding and H.~Peng, ``Minimum redundancy feature selection from microarray
  gene expression data,'' \emph{Journal of bioinformatics and computational
  biology}, vol.~3, no.~02, pp. 185--205, 2005.

\bibitem{zhang2015accelerometer}
Y.~Zhang, G.~Pan, K.~Jia, M.~Lu, Y.~Wang, and Z.~Wu, ``Accelerometer-based gait
  recognition by sparse representation of signature points with clusters,''
  \emph{IEEE Transactions on Cybernetics}, vol.~45, no.~9, pp. 1864--1875,
  2015.

\bibitem{wei2013real}
B.~Wei, M.~Yang, Y.~Shen, R.~Rana, C.~T. Chou, and W.~Hu, ``Real-time
  classification via sparse representation in acoustic sensor networks,'' in
  \emph{Proceedings of the ACM Conference on Embedded Networked Sensor
  Systems}, 2013.

\bibitem{weitao2016sensor}
W.~Xu, Y.~Shen, N.~Bergmann, and W.~Hu, ``Sensor-assisted face recognition
  system on smart glass via multi-view sparse representation classification,''
  in \emph{Proceedings of the ACM/IEEE International Conference on Information
  Processing in Sensor Networks}, 2016.

\bibitem{aharon2006img}
M.~Aharon, M.~Elad, and A.~Bruckstein, ``K-svd: An algorithm for designing
  overcomplete dictionaries for sparse representation,'' \emph{IEEE
  Transactions on Signal Processing}, vol.~54, no.~11, pp. 4311--4322, 2006.

\bibitem{engan2000multi}
K.~Engan, S.~O. Aase, and J.~H. Hus{\o}y, ``Multi-frame compression: Theory and
  design,'' \emph{Signal Processing}, vol.~80, no.~10, pp. 2121--2140, 2000.

\bibitem{lee2001algorithms}
D.~D. Lee and H.~S. Seung, ``Algorithms for non-negative matrix
  factorization,'' in \emph{Proceedings of Advances in Neural Information
  Processing Systems}, 2001.

\bibitem{WrightYangGaneshSastryMa09}
J.~Wright, A.~Yang, A.~Ganesh, S.~Sastry, and Y.~Ma, ``{Robust face recognition
  via sparse representation},'' \emph{IEEE Transactions on Pattern Analysis and
  Machine Intelligence}, vol.~31, no.~2, pp. 210--227, 2009.

\bibitem{natarajan1995sparse}
B.~K. Natarajan, ``Sparse approximate solutions to linear systems,'' \emph{SIAM
  Journal on Computing}, vol.~24, no.~2, pp. 227--234, 1995.

\bibitem{CandesRombergTao06}
E.~J. Candes, J.~Romberg, and T.~Tao, ``{Robust uncertainty principles: exact
  signal reconstruction from highly incomplete frequency information},''
  \emph{IEEE Transactions on Information Theory}, pp. 489--509, 2006.

\bibitem{Donoho06}
D.~Donoho, ``{Compressed sensing},'' \emph{IEEE Transactions on Information
  Theory}, pp. 1289--1306, 2006.

\bibitem{MIDE}
``Mide technology,'' \url{http://www.mide.com/}.

\bibitem{shin2015urban}
D.~Shin, D.~Aliaga, B.~Tun{\c{c}}er, S.~M. Arisona, S.~Kim, D.~Z{\"u}nd, and
  G.~Schmitt, ``Urban sensing: Using smartphones for transportation mode
  classification,'' \emph{Computers, Environment and Urban Systems}, vol.~53,
  pp. 76--86, 2015.

\bibitem{liu2013ambient}
V.~Liu, A.~Parks, V.~Talla, S.~Gollakota, D.~Wetherall, and J.~R. Smith,
  ``Ambient backscatter: Wireless communication out of thin air,'' \emph{ACM
  SIGCOMM Computer Communication Review}, vol.~43, no.~4, pp. 39--50, 2013.

\bibitem{aktakka2015three}
E.~E. Aktakka and K.~Najafi, ``Three-axis piezoelectric vibration energy
  harvester,'' in \emph{Proceedings of the IEEE International Conference on
  Micro Electro Mechanical Systems}, 2015.

\bibitem{hung2015miniature}
C.-F. Hung, T.-K. Chung, P.-C. Yeh, C.-C. Chen, C.-M. Wang, and S.-H. Lin, ``A
  miniature mechanical-piezoelectric-configured three-axis vibrational energy
  harvester,'' \emph{IEEE Sensors Journal}, vol.~15, no.~10, pp. 5601--5615,
  2015.

\bibitem{lecun1998gradient}
Y.~LeCun, L.~Bottou, Y.~Bengio, P.~Haffner \emph{et~al.}, ``Gradient-based
  learning applied to document recognition,'' \emph{Proceedings of the IEEE},
  vol.~86, no.~11, pp. 2278--2324, 1998.

\bibitem{zhou2016learning}
B.~Zhou, A.~Khosla, A.~Lapedriza, A.~Oliva, and A.~Torralba, ``Learning deep
  features for discriminative localization,'' in \emph{Proceedings of the IEEE
  Conference on Computer Vision and Pattern Recognition}, 2016.

\bibitem{xiao2016learning}
T.~Xiao, H.~Li, W.~Ouyang, and X.~Wang, ``Learning deep feature representations
  with domain guided dropout for person re-identification,'' in
  \emph{Proceedings of the IEEE Conference on Computer Vision and Pattern
  Recognition}, 2016.

\bibitem{yosinski2014transferable}
J.~Yosinski, J.~Clune, Y.~Bengio, and H.~Lipson, ``How transferable are
  features in deep neural networks?'' in \emph{Proceedings of the Advances in
  Neural Information Processing Systems}, 2014.

\bibitem{xie2016transfer}
M.~Xie, N.~Jean, M.~Burke, D.~Lobell, and S.~Ermon, ``Transfer learning from
  deep features for remote sensing and poverty mapping,'' in \emph{Proceedings
  of the {AAAI} Conference on Artificial Intelligence}, 2016.

\end{thebibliography}
\balance
\end{document}